%% file: BeamNeutronResult2018Data.tex
\documentclass[a4paper,11pt]{article}
\pdfoutput=1
\usepackage{jinstpub}

\usepackage[mathlines]{lineno}
\usepackage{siunitx} 
\usepackage{xspace}
\usepackage{csquotes}
\usepackage{graphbox}

\begin{document}
\title{Monitoring the SNS basement neutron background with the MARS detector}

\input{authors/authors-MARS2018.tex}

\emailAdd{bcabrer@sandia.gov}

\abstract{We present the analysis and results of the first dataset 
collected with the MARS neutron detector 
deployed at the Oak Ridge National 
Laboratory Spallation Neutron Source (SNS) for the purpose of 
monitoring and characterizing the beam-related neutron (BRN) background 
for the COHERENT collaboration. MARS was positioned 
next to the COH-CsI coherent elastic neutrino-nucleus scattering detector
in the SNS basement corridor. This is the basement location of
closest proximity to the SNS target and thus, of highest neutrino flux,
but it is also well shielded from the BRN flux by infill concrete 
and gravel. These data show the detector registered roughly one BRN per day. 
Using MARS' measured detection efficiency, the incoming 
BRN flux is estimated to be $1.20~\pm~0.56~\text{neutrons}/\text{m}^2/\text{MWh}$ 
for neutron energies above $\sim3.5$~MeV and up to a few tens of MeV. 
We compare our results with previous BRN measurements in the SNS basement corridor
reported by other neutron detectors.}


\arxivnumber{}

\maketitle

\section{\label{sec:level1}Introduction}

Understanding the fast and higher-energy ($\gtrapprox10$~MeV) neutron background 
is of particular importance for experiments aiming to measure coherent elastic 
neutrino-nucleus scattering (CEvNS) since neutron elastic scattering produces
 nuclear recoils similar to the 
CEvNS signature. For the COHERENT experiment 
~\cite{COHERENT:2017ipa, COHERENT:2018imc, COHERENT:2018gft, COHERENT:2019iyj,
 COHERENT:2020iec,COHERENT:2020ybo, COHERENT:2021xhx,COHERENT:2019kwz} 
at the Oak Ridge National 
Laboratory Spallation Neutron Source (SNS), the most impactful 
neutrons are those produced in the SNS target at the same time as the neutrinos, 
the latter originating also at the target from the decay of pions.

The 1-GeV, 1.4-MW proton beam of the SNS accelerator strikes a liquid-Hg 
target in 360 ns FWHM pulses at 60 Hz to produce neutrons that are moderated 
and delivered to neutron experiments. As part of each proton-on-target 
(POT) event, prompt muon neutrinos ($\nu_{\mu}$) and delayed muon antineutrinos 
and electron neutrinos  ($\bar{\nu}_{\mu}$ and $\nu_{e}$) are also produced 
from the decay of pions~\cite{COHERENT:2018gft,COHERENT:2021xhx}, 
benefiting neutrino-physics experiments like COHERENT.
To mitigate the impact of the beam-related neutron 
(BRN) background, the COHERENT detectors are deployed in the SNS basement 
corridor dubbed \enquote{Neutrino Alley}. 
This location is shielded from the SNS target by the combination 
of the target monolith and infill concrete and gravel, 
while it also provides the benefit 
of 8 meter of water equivalent overburden against 
cosmic rays~\cite{collaboration2015coherent}. Nevertheless, knowledge of 
the rate and spectral distribution of the beam neutrons reaching the 
various COHERENT detectors is important for detector design, modeling and 
data analysis. For the CEvNS measurement with the COH-CsI detector
\cite{COHERENT:2017ipa} deployed at the location with the lowest BRN flux according to 
previous measurements~\cite{collaboration2015coherent}, the BRN background 
was about 6\% of the predicted CEvNS rate, with a 25\% systematic uncertainty. 
However, for the CEvNS observation with argon~\cite{COHERENT:2019iyj}, 
the BRN rate was about four times larger than the predicted CEvNS rate.
Developing a firm understanding of the neutron flux in Neutrino Alley is 
thus critical for future high-precision cross-section measurements.

In this paper, we provide the analysis and results of neutron data collected 
with the Multiplicity and Recoil Spectrometer, or MARS, 
deployed at the SNS for the purpose of monitoring and characterizing the 
BRN background for the COHERENT collaboration. As will be described in 
section~\ref{s:detector}, the MARS configuration deployed at the SNS and 
discussed throughout this paper includes 
one module of plastic scintillator layers with interleaved gadolinium (Gd) coated 
Mylar sheets, and operates in the \enquote{capture-gated} mode explained in 
section~\ref{ss:detection}. This differs from the original configuration of 
this detector created for a different experiment~\cite{ROECKER201621}, which 
also included a plane of lead bricks sandwiched between two modules 
of plastic scintillator layers and exploited the \enquote{multiplication} of 
neutrons in the lead bricks due to incoming high-energy neutrons.

Details of the MARS experimental deployment at the SNS are covered in 
section~\ref{s:data}. Sections~\ref{s:processing} and \ref{s:analysis}
are devoted to the processing and analysis of the first beam-on MARS 
dataset collected during the year 2018. The measured BRN rates at the 
MARS deployment location beside the COH-CsI neutrino detector 
of~\cite{COHERENT:2017ipa} are reported in section~\ref{s:alleyBRNflux}. 
A comprehensive study of the MARS response--- including the 
measurement and modeling of its light-collection efficiency, 
energy resolution and capture-gated signal response 
as a function of active volume region, the derivation of the 
energy calibration as well as the modeling of 
the detection-efficiency energy and threshold dependence--- will be the subject of 
future work. In this work, for the purpose of providing an estimate 
of the incident BRN flux, section~\ref{s:alleyBRNflux}
contains a summary of MARS' detection efficiency measurement done 
with 14~MeV neutrons. MARS data taken subsequently at different 
Neutrino Alley locations with different BRN flux levels 
will be treated in future work. 


\section{\label{s:detector}MARS detector for BRN background studies at SNS}

\subsection{\label{ss:detection}Neutron-detection concept}
For MARS' neutron-detection mechanism, we use the so-called capture-gated 
mode. When a fast neutron elastically interacts with hydrogen nuclei in 
MARS' large plastic scintillator volume, the recoiling protons generate a prompt 
scintillation pulse. If the neutron does not escape, it will bounce 
around the detector while quickly thermalizing. In the presence of a 
material with a large thermal-neutron capture cross section like Gd, 
there is a high probability that de-excitation gamma rays will be produced. 
For Gd, the maximum total energy of the emitted gamma rays is \(\sim 8\)~MeV.
The neutron-capture process in MARS has a time constant \(\tau_{\text{nCapt}}\) 
that depends on the Gd concentration, and it was reported to be \(18.7~\mu s\)
for the MARS configuration used in~\cite{ROECKER201621}.
Thus, by employing the timing and energy of the Gd gamma-ray scintillation 
pulses, the population of fully thermalized and absorbed neutrons can be 
isolated. When the neutron's kinetic energy is fully and quickly 
transferred to the scintillator in elastic scatterings, the integral 
of the prompt pulse is a direct measure of the incoming neutron energy.
Since the fraction of neutrons thermalizing purely via elastic interactions 
decreases with neutron energy, being of the order of a few percent for 
\(\sim 50\)~MeV neutrons,
the capture-gated mode as considered in this work is limited to provide 
rate and spectral information on BRN energies no larger than a few tens of MeV. 

\subsection{\label{ss:hardware}Hardware description}

\begin{figure}
\centering
\includegraphics[width=0.40\textwidth]{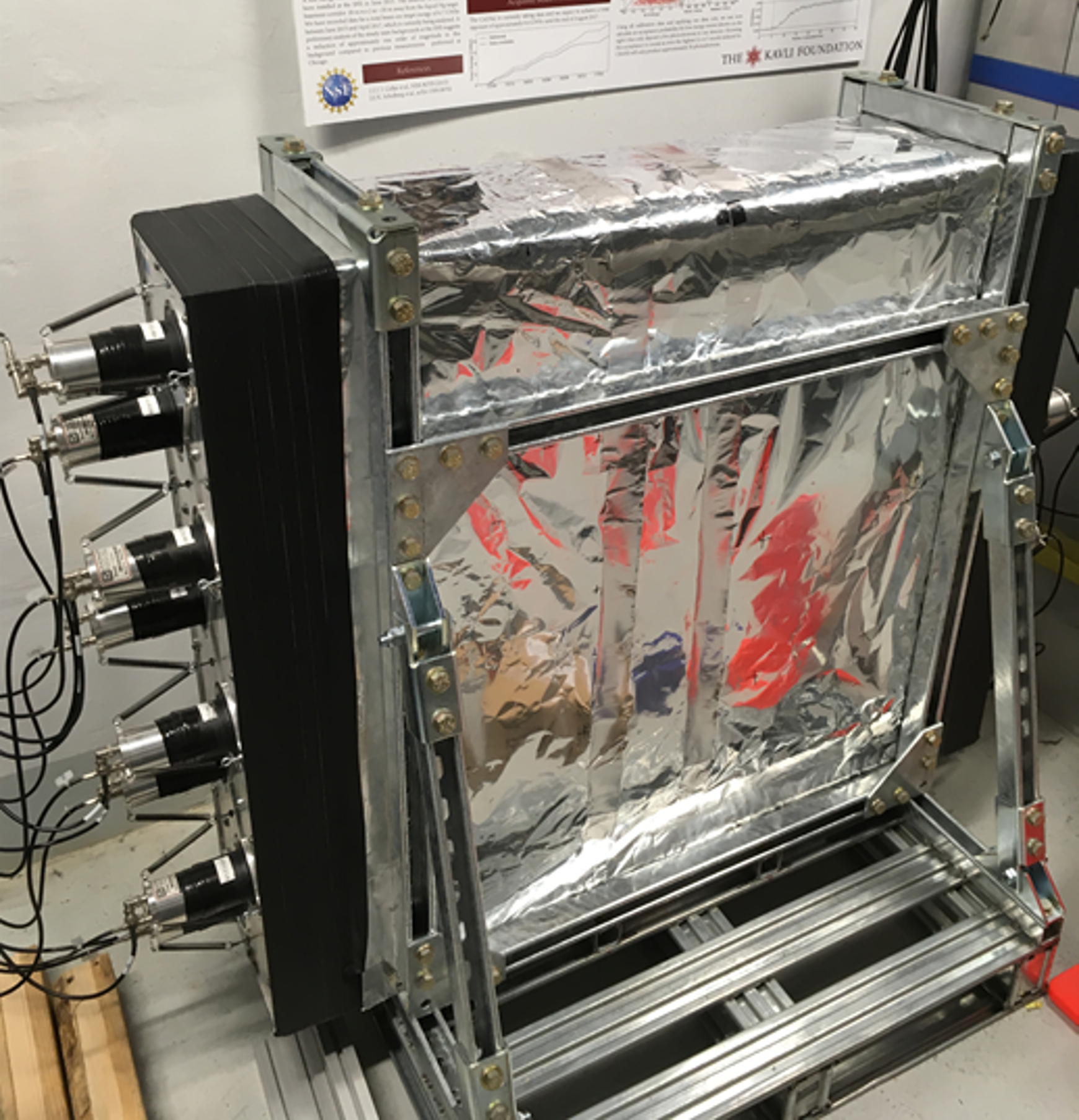}
\hspace{1cm}
\includegraphics[width=0.30\textwidth]{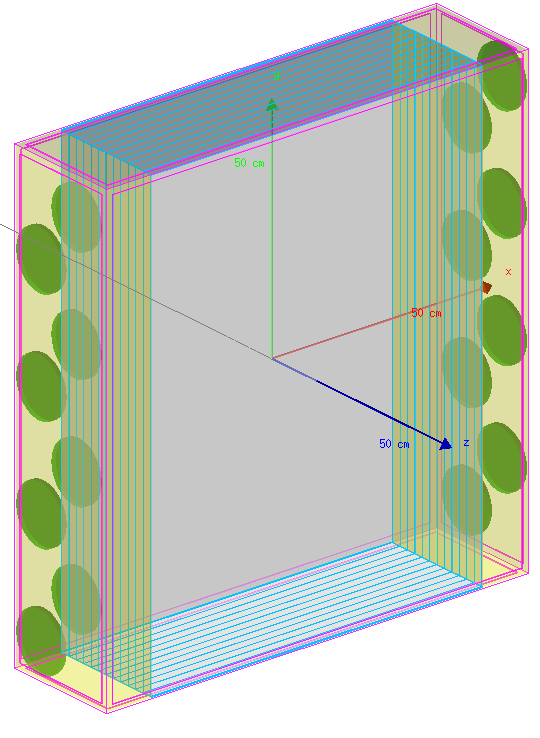}
\caption{Left: A photograph of the MARS detector in Neutrino Alley. 
The bases of the PMTs are visible on the left-hand side of the photograph, 
and a unistrut support structure holds the stacked scintillators.
Right: Geant4~\cite{geant4} model of the detector. The cyan boxes represent the 
plastic scintillator layers. The side acrylic light guides are in yellow, 
with green circles representing the PMT windows.} 
\label{f:MARSdetector}
\end{figure}

The MARS detector, pictured in Fig.~\ref{f:MARSdetector} (left), 
consists of twelve 2-cm-thick BC-408 plastic scintillator layers 
interleaved with Gd-coated Mylar sheets. The overall dimensions 
of the detection volume are \(L\times W\times H = 75\times 25\times 100~\text{cm}^3\). 
This module stands vertically on its \(L\times W\) side held by a 
Unistrut frame, with each \(W\times H\) face covered by a 
\(10\times 25\times 100~\text{cm}^3\) acrylic light guide
that optically couples the scintillator layers. 
On each side, 
eight 5-inch diameter ADIT B133D01 photomultiplier tubes (PMTs) are 
coupled to the light guides by silicon grease. Black tape covers the 
scintillator and acrylic volumes for light-tightness, which are further 
wrapped with an aluminum sheet for fire safety. 

\subsection{\label{ss:electronics}Electronic read-out and trigger scheme}

\begin{figure}
\centering
\includegraphics[width=0.70\textwidth]{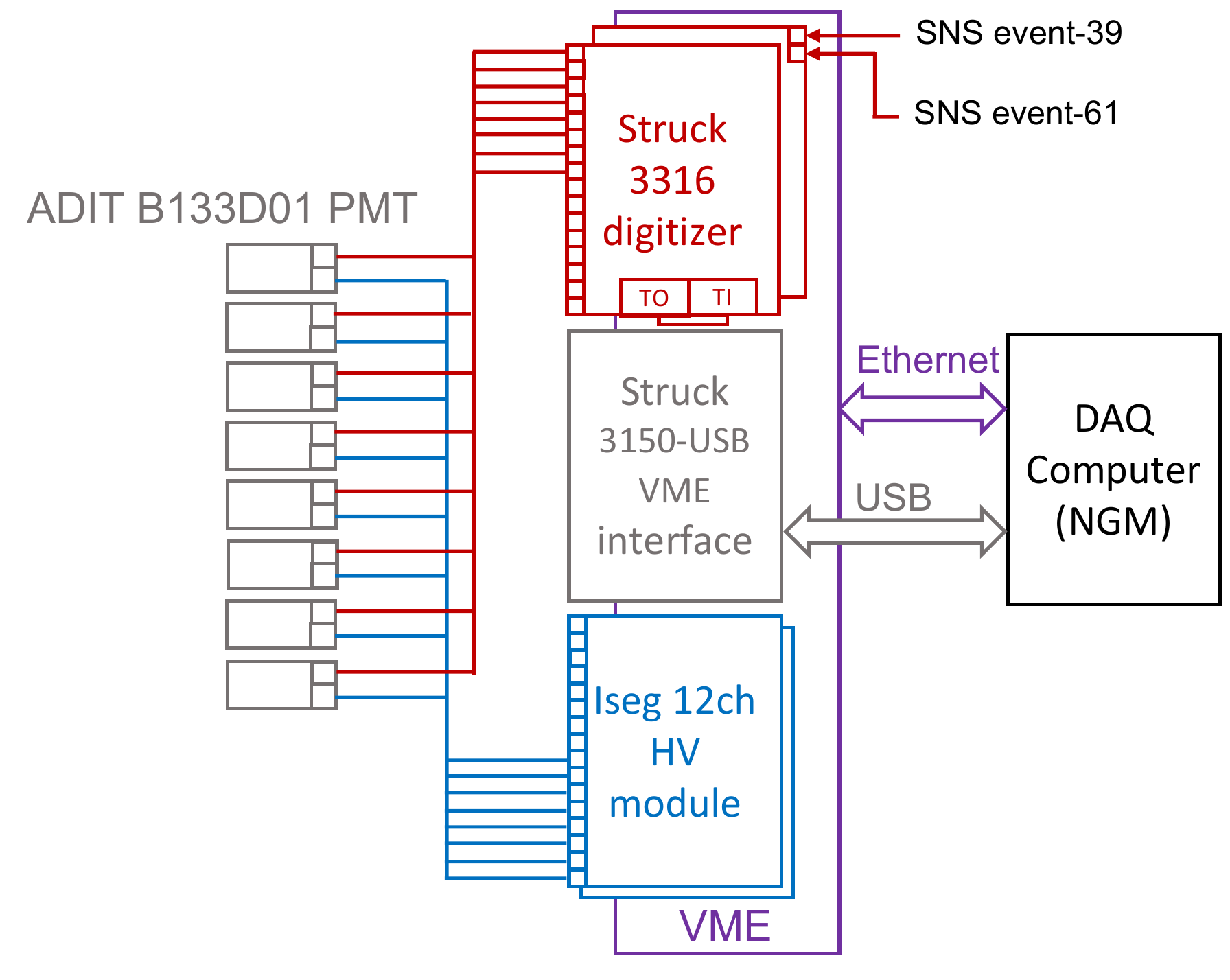}
\caption{Diagram of MARS's electronic read-out and high-voltage (HV) setup. 
Only eight of the 16 PMT channels are represented for simplicity. The data acquisition 
(DAQ) is managed by the NGM software package~\cite{NGM}. 
TO and TI represent the Trigger Out and 
Trigger In connectors, respectively.} 
\label{f:daq_diagram}
\end{figure}

Data readout, diagrammatically shown in Fig.~\ref{f:daq_diagram}, is done 
with two 14-bit, 250 MHz Struck SIS3316 sixteen-channel 
waveform digitizers~\cite{Struck} mounted in a VME crate.  All channels of the 
first digitizer are used to digitize the sixteen PMT signals. The second 
digitizer asynchronously records two SNS-provided signals,
arbitrarily named \enquote{event-39} and \enquote{event-61} triggers, 
which are synchronized with the SNS beam. The event-39 triggers are 
provided uninterrupted at 60 Hz during beam-on and beam-off periods. 
The event-61 triggers are supplied during beam-on operation whenever 
there is a POT event, which occurs also with a 60 Hz frequency  
but with one event skipped in every (1/600) second period.


The signals from a pair of neighboring PMTs on one side of the scintillator module 
and the corresponding pair at the same height on the opposite side 
are joined into the same digitizer's channel group, for a total of 
four groups. Using the digitizer's trigger logic, a trapezoidal Finite 
Input Response (FIR) filter is applied on each four-channel sum signal 
to generate a trigger signal~\cite{Struck}. The FIR filter's peaking and 
gap times were each set to 4 samples, or 16 ns. With the CFD (Constant Fraction 
Discriminator) feature enabled, the trapezoidal threshold value 
is set to achieve the lowest physical signal threshold that still produces 
negligible dead time --- the physical signal threshold in energy units 
will be discussed in section~\ref{s:processing}. The internally 
generated trigger is routed to the external trigger input to start 
the sampling of all sixteen channels. 
The pre-trigger delay was adjusted to contain most of the scintillation pulse
 within the 150-sample active trigger gate window, including 
50 baseline pre-samples.

For each recorded event, the raw channel data contain: channel ID, 
timestamp, peak sample index, peak height value, and six accumulator sums. 
Each accumulator stores the integral of 25 samples, with accumulator 0 
integrating samples 0-24, accumulator 1 integrating samples 25-49, and so on. 
At the first stage of processing, each PMT channel integral is calculated as the 
sum of accumulators 2 to 5 (equivalent to integrating 100 samples) minus the 
baseline integral. The latter is computed as the sample average of the combined 
first two accumulators multiplied by 100. The sum of the sixteen baseline-subtracted 
integrals is taken to represent the total energy of the scintillator pulse in
the digitizer output units which we denote here as \enquote{adc} units. 
The time intervals since the previous event-39 signal, the previous event-61 signal and 
the previous scintillator pulse are also recorded.  A further description of the 
pre-processed data is given at the start of section~\ref{s:processing}.

\section{\label{s:data}Experimental data}

\subsection{\label{s:location}MARS location in the SNS basement corridor}

\begin{figure}
\centering
\includegraphics[width=0.70\textwidth]{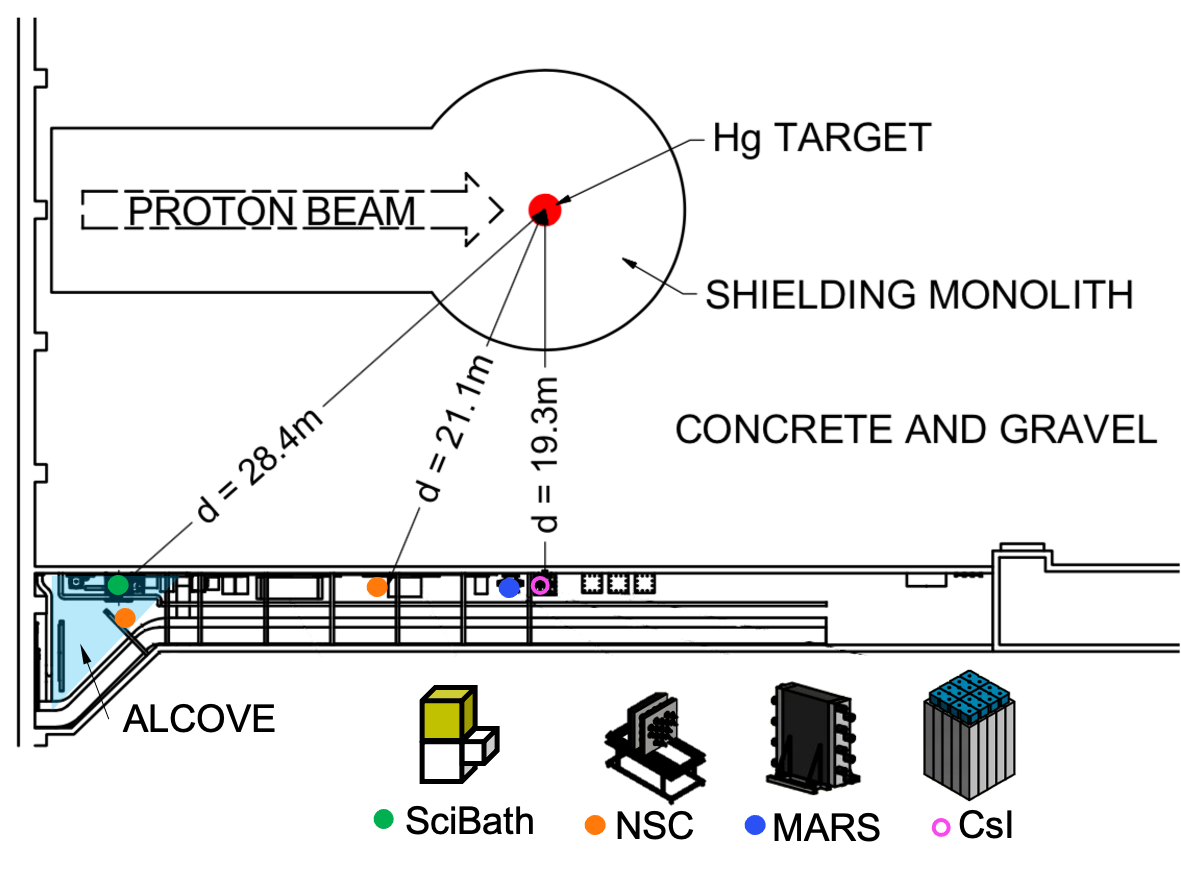}
\caption{Diagram of a top-down view of the SNS Neutrino Alley showing 
 MARS and CsI detector locations during 2018. Also shown are the earlier deployment 
locations of the neutron scatter camera (NSC)~\cite{CabreraPalmer2019} and of
the SciBath detector~\cite{Brice:2013fwa}, employed as part of 
COHERENT's first neutron-background measurements at the SNS. The blue shade 
indicates the Neutrino Alley alcove area. Some of COHERENT's other detectors 
are omitted here for simplicity.} 
\label{f:neutrinoAlley}
\end{figure}

In the fall of 2017, MARS was deployed in the SNS basement corridor next to the 
COH-CsI detector system (Fig.~\ref{f:neutrinoAlley}, right), at 19.5~m from the SNS 
target, with concrete and gravel filling most of the space between the Neutrino 
Alley and the SNS target monolith, plus 8 meter-water-equivalent overburden. 
A previous neutron-background measurement campaign~\cite{collaboration2015coherent} 
that included three different locations along the SNS basement corridor showed the vicinity 
of the COH-CsI system as having the lowest SNS BRN background. In this regard, 
the BRN rates measured by MARS here are expected to be highly suppressed. 

A high flux of 511~keV gamma rays is present all along the Neutrino Alley. 
This gamma field is emitted from a 6-inch diameter pipe that runs through the 
corridor near the concrete ceiling and carries radioactive gas from 
the SNS-target water moderating system. These 511~keV gamma rays 
mainly originate from $\beta^{+}$~decay positrons (\textit{e.g.}, from $^{11}\text{C}$)
and create a high and continually 
present \enquote{Hot Off Gas} (HOG) background during beam-on periods. 
Given that the MARS design does not include shielding material of any kind, 
its response has been found to be highly sensitive to variations in the 
HOG background during the analyzed period, as will be presented in 
section~\ref{s:processing}. 
Adding 1 cm of high-Z shielding material like lead around the MARS active volume
would significantly attenuate the 511~keV gamma flux, but it would also create a
source of multiple low energy neutrons ($\sim$1-2 MeV, \cite{ROECKER201621,roecker_2016}) 
from the spallation reactions of the high-energy BRNs. 
Without a careful quantization of this  effect, MARS was left without
shielding, and instead, efforts have focused on
surrounding the HOG pipe with a lead layer to suppress the 511~keV background. 

\subsection{\label{s:period}Data-collection period and data quality}

\begin{figure}
\centering
\includegraphics[width=0.95\textwidth]{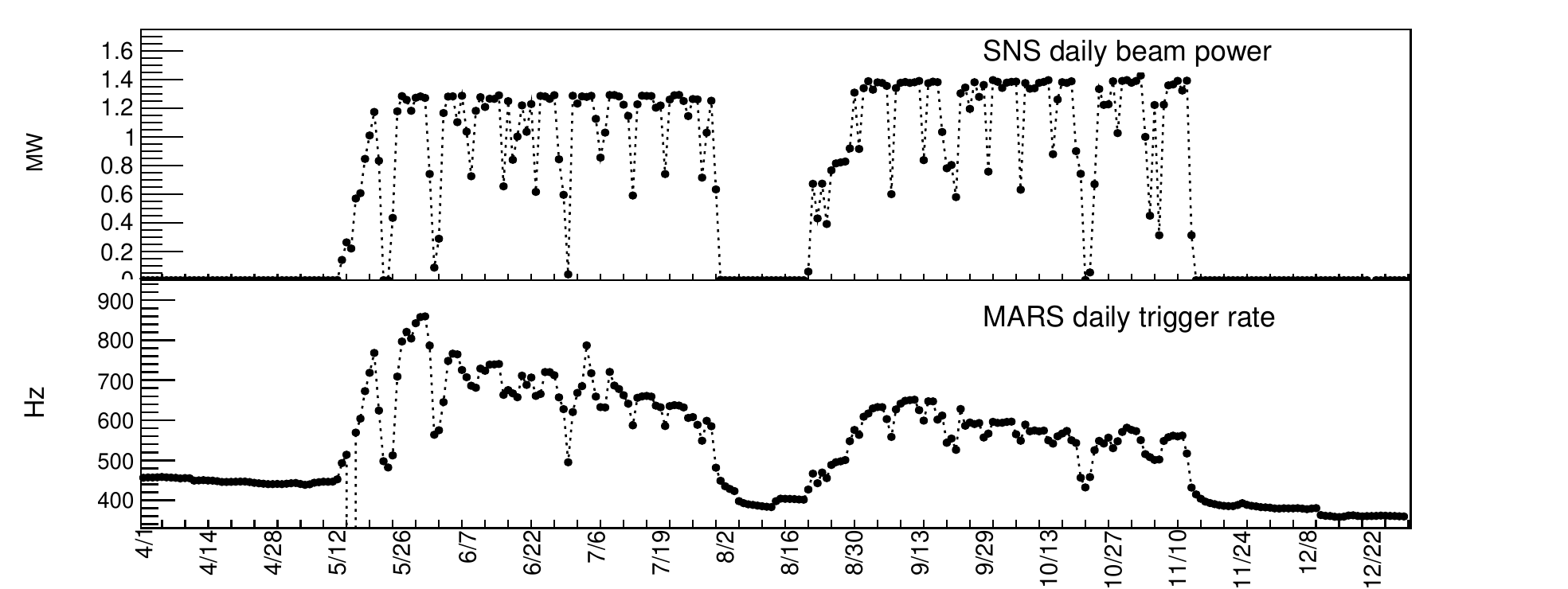}
\caption{ 
Top: SNS daily delivered beam power. Bottom:
daily PMT-sum trigger rate as reported by the COHERENT 
Grafana~\cite{Web:Grafana:Docs} dashboard.
The horizontal axis represents the calendar dates 
from April 1, 2018 to December 31, 2018, and are 
labeled as month/day. The PMT-rate gradual decrease 
indicates a gain shift in the MARS response.}
\label{f:beampower_rates_sizes}
\end{figure}

MARS production data collection started at the end of December 2017 during an  
SNS shutdown period, so that sufficient beam-off data were collected before 
the SNS beam turned on in May 2018. The analysis presented here
uses data from April 01, 2018 to December 31, 2018. The top panel of 
Fig.~\ref{f:beampower_rates_sizes} shows the beam power for each day of the period under analysis. 
The SNS operational beam power level intended 
for the beam-on period covering June and July was \(1.3\)~megawatt (MW), 
increasing to \(1.4\)~MW for the subsequent beam-on period 
from September to early November. 

The PMT trigger rates, presented for each day in the second 
panel of Fig.~\ref{f:beampower_rates_sizes}, show a significant 
jump during the two 2018 beam-on periods compared to the beam-off data, 
due to the intense HOG gamma-background flux. 
The 2018 PMT rates also showed a gradual decrease during beam-on periods when
the beam power was stable --- and thus, with the HOG background intensity 
also expected to be stable --- indicating 
a gain shift in the MARS response. We show in section~\ref{ss:muonN_selection} 
how these 511~keV background signals not only increase the total event rate but 
also pile-up on most of the PMT signal pulses, causing a positive shift in the pulse-integral 
energy spectrum. Thus, in order to elude the HOG-background shifting effect on the spectrum,
we compared the April, August and December beam-off data and noted that the event rate 
always decreased relative to the previous beam-off period. Since the digitizer 
threshold settings were kept unchanged during the whole data-collection period, 
we hypothesize that the event-rate decrease arose 
from an effective decrease in overall gain over time, though its origin 
was not investigated. 

During year 2018 production operation, a new run 
was started every 24 hours. The second panel of Fig.~\ref{f:beampower_rates_sizes}
shows the data's daily trigger rate,
illustrating that there were not any major data-collection interruptions. 
In our analysis, the data were aggregated in 14-day intervals. Most of the intervals 
correspond to sets of continuous 14 days; in the cases when there was a calendar-day gap, 
the next calendar day at the end of the interval was aggregated in order to keep the 
total aggregated time  to be approximately 14 days. 
 

\section{\label{s:processing}Data Processing}

In data pre-processing, any two scintillator pulses separated by no more 
than \(200~\mu s\)  are grouped into a \enquote{pulse pair} --- for example, 
a sequence of 3 scintillation pulses where the first and last were closer 
than \(200~\mu s\) would yield 3 pulse pairs. 
Each pair is described by the variables \( (t_1, E_1, \Delta t, E_2) \), where \( t_1 \) 
is the time interval between the first pulse of the pair, or prompt pulse, 
and the preceding event-39 trigger (and thus, \( t_1 < 1/60\)~s), \( E_1 \) 
represents the first pulse's energy in \enquote{adc} units,  
\( \Delta t \) is the inter-pulse time (restricted to be \( \Delta t < 200~\mu s\)) and 
\( E_2 \) is the energy of the second pulse, or delayed 
pulse, also in \enquote{adc} units. When a capture-gated neutron detection occurs, \( (t_1, E_1) \) 
represent the prompt 
neutron-pulse time and energy, while 
\( (\Delta t, E_2) \) represent the Gd neutron-capture time and the energy of 
the Gd-decay gamma-ray pulse. The \(200~\mu s\) maximum bound on \( \Delta t\) 
will miss a negligible number of neutron-capture events since it is 
more than 10 times the previously measured Gd neutron-capture time constant. 

\begin{figure}
\centering
\includegraphics[width=0.49\textwidth]{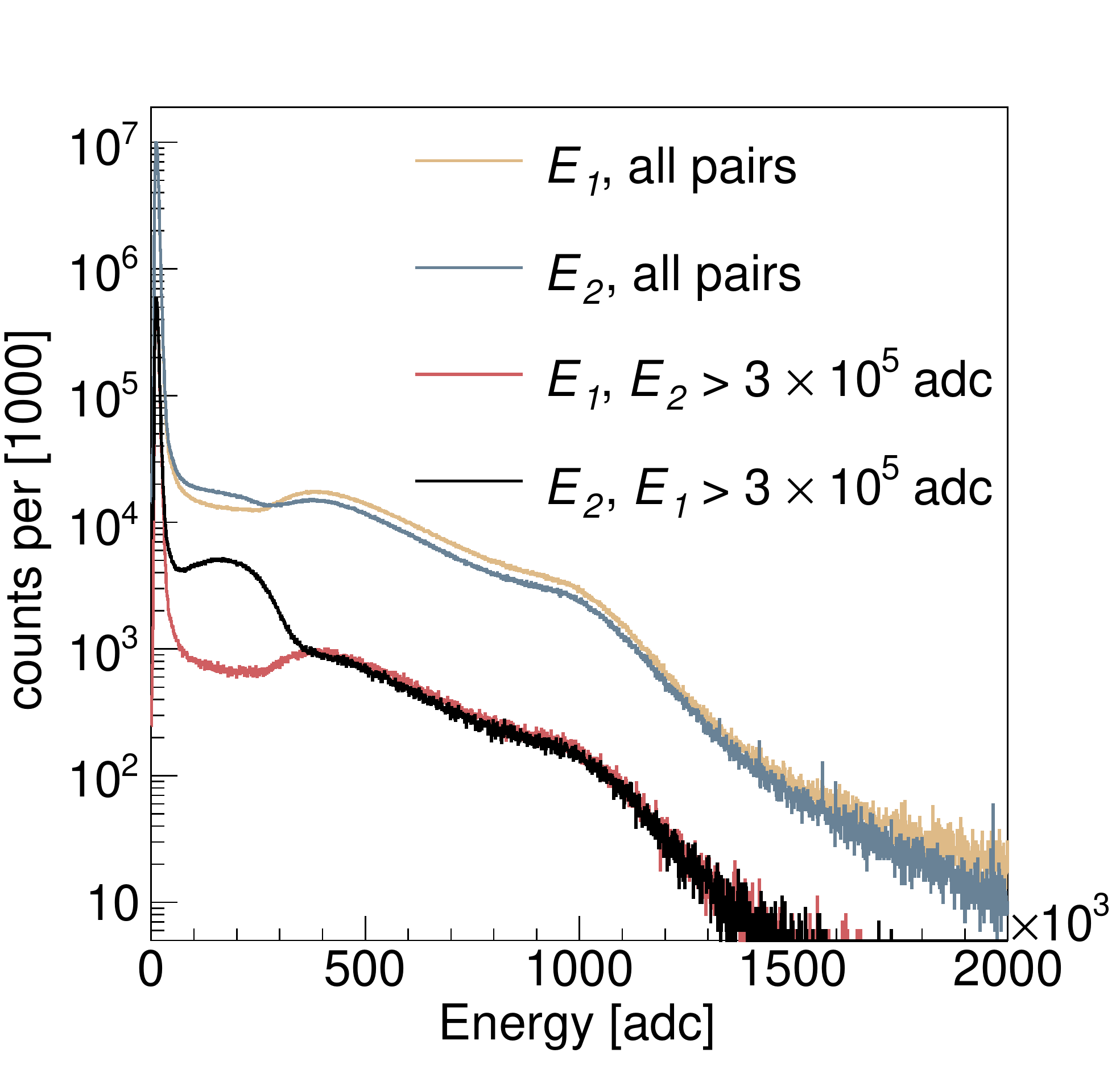}
\includegraphics[width=0.49\textwidth]{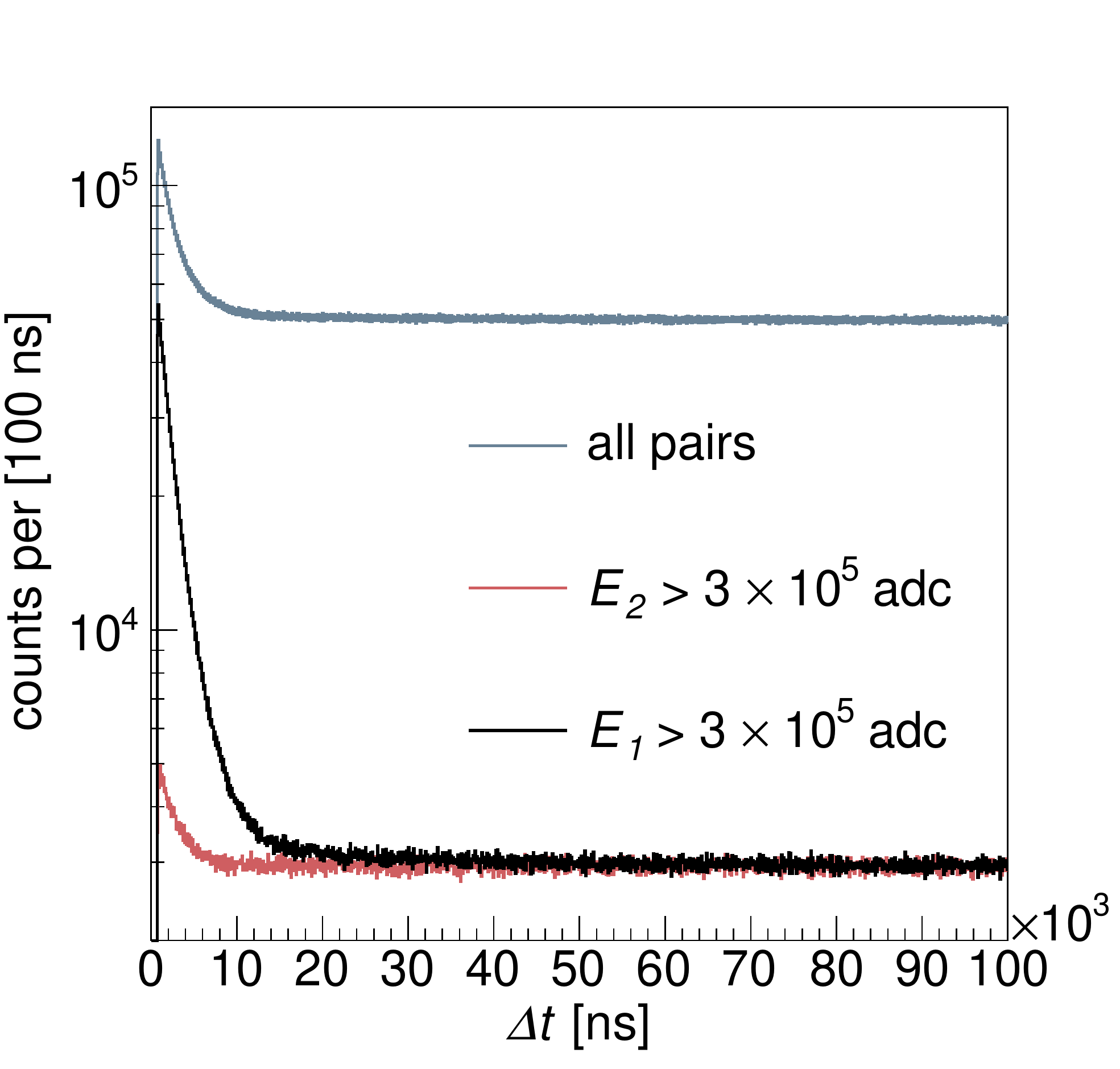}
\caption{Left: Histograms of the first pulse's energy \(E_1\) and 
the second pulse's energy \(E_2\) of the scintillator pulse pairs 
from the April 2018 beam-off data. 
The first two histograms in the legend include all pairs. The last (black) 
\(E_2\) histogram only includes pairs with \(E_1 > 3 \times 10^{5}\)~adc 
that select muon-track high-energy depositions. The (red) \(E_1\) histogram, 
with the same cut applied instead to \(E_2\), illustrates the differences 
in the \(E_1\) and \(E_2\) populations. Right: Histograms of the inter-pulse 
time \(\Delta t\) for the same cuts. The black histograms of both panels clearly show the spectral 
(\(E_2\)) and time (\(\Delta t\)) profiles expected for Michel electrons.}
\label{f:muonE1E2dt_April2018}
\end{figure}

\begin{figure}
\centering
\includegraphics[width=0.49\textwidth]{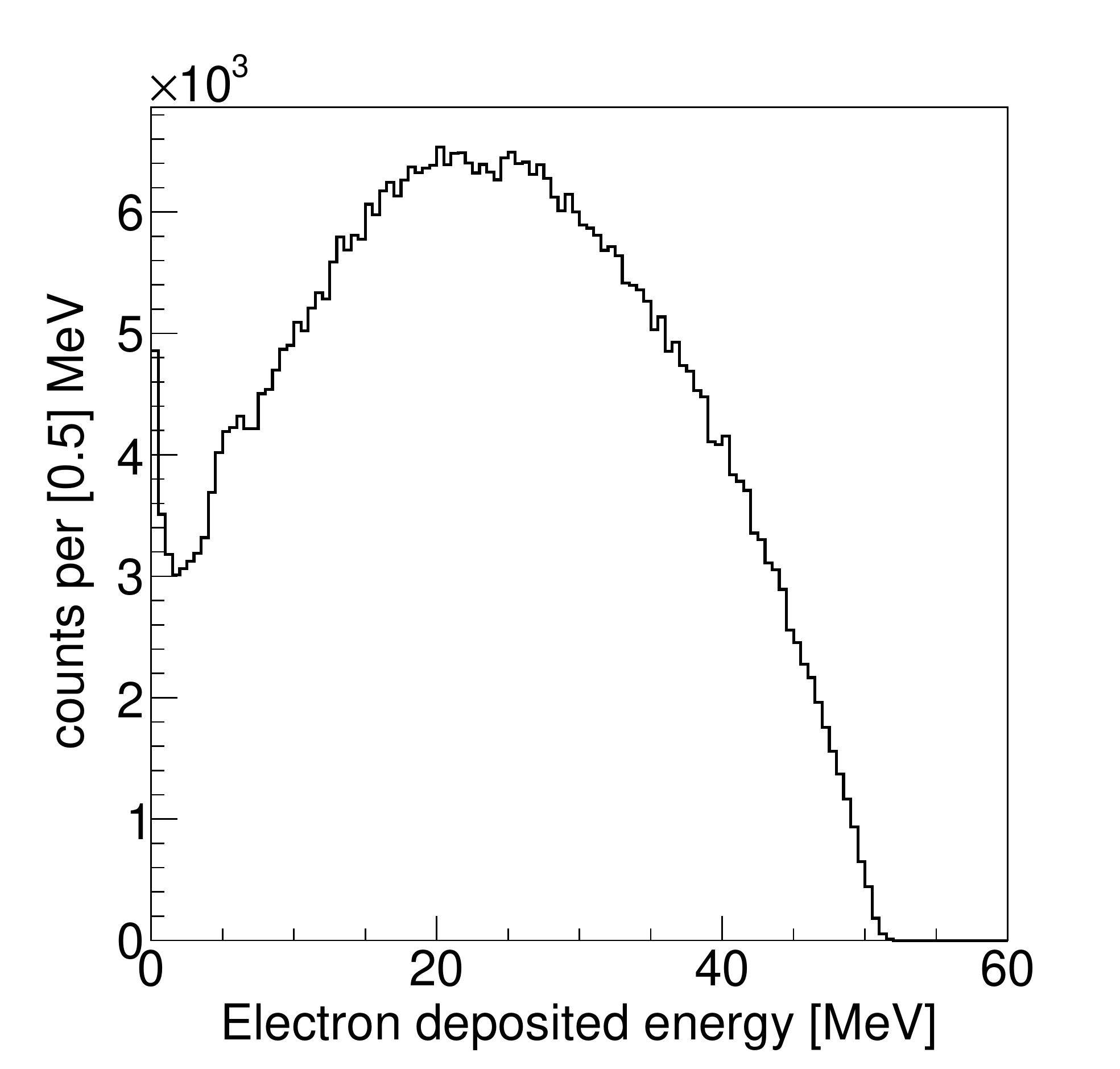}
\caption{Geant4 simulation of Michel electron tracks that are randomly started within MARS 
scintillator volume. The simulated electron deposited energy histogram shows the 
round peak at \(\sim 20\)~MeV. 
Since no energy smearing and light-collection non-uniformity effects have been included, 
the deposited energy units are expressed in MeV, instead of the MeVee units used for 
the detected light output.  }
\label{f:michel_e_sim}
\end{figure}

The derivation of the energy calibration to 
convert from \enquote{adc} units to the energy deposit expressed in 
scintillation light output units \enquote{MeV electron equivalent}, 
or MeVee, is out of the scope of this paper. 
For the purpose of assessing  
the consistency of this paper's results, the scaling $\sim 10^4$~adc per 1~MeVee
provides a crude approximation of the energy calibration. This scaling places
our trigger threshold at \(\sim 1\)~MeVee (\(10^4\) adc), 
or \(\sim 3.5\)~MeV in proton recoil energy~\cite{Jian_Fu_2010,Laplace:2020mfy}.
As an illustration, 
Fig.~\ref{f:muonE1E2dt_April2018} shows the $E_1$, $E_2$ and $\Delta t$ 
histograms of all pairs for one beam-off month. Cosmic muons traversing MARS 
can generate large pulses and represent the main contribution to the 
high-energy end of the $E_1$ and $E_2$ spectra. Muons decaying within 
the detector produce highly energetic Michel electrons with a decay time constant 
\(\tau_{\mu -}= 2.2~\mu s\)~\cite{Agashe:2014kda} and a kinetic-energy cutoff 
at \(\sim\)50~MeV~\cite{Acciarri:2017sjy}. These events are included in our 
selected data since they also have a double-hit \((E_1,E_2)\) 
signature corresponding to the muon's ionization \(E_1\) as the muon 
penetrates the detection medium and stops, followed by the Michel electron deposition 
\(E_2\). The black histogram of the left panel of Fig.~\ref{f:muonE1E2dt_April2018} 
shows the prominent and wide Michel electron peak at \(E_2\sim 20 \times 10^{4}\)~adc,
 obtained by restricting \(E_1 > 30 \times 10^{4}\)~adc 
to select muon-track high-energy depositions. The \(\Delta t\) profile of these 
pairs, presented in the right panel of Fig.~\ref{f:muonE1E2dt_April2018}, 
exhibits the Michel electron timing profile. 
In Fig.~\ref{f:michel_e_sim}, the Geant4~\cite{geant4} simulation of 
Michel electron tracks started at random locations within the scintillator volume 
shows their ionization energy spectrum peaking near \(\sim 20\)~MeV, consistent with
 our crude energy scaling. 

\begin{figure}
\includegraphics[width=0.49\textwidth]{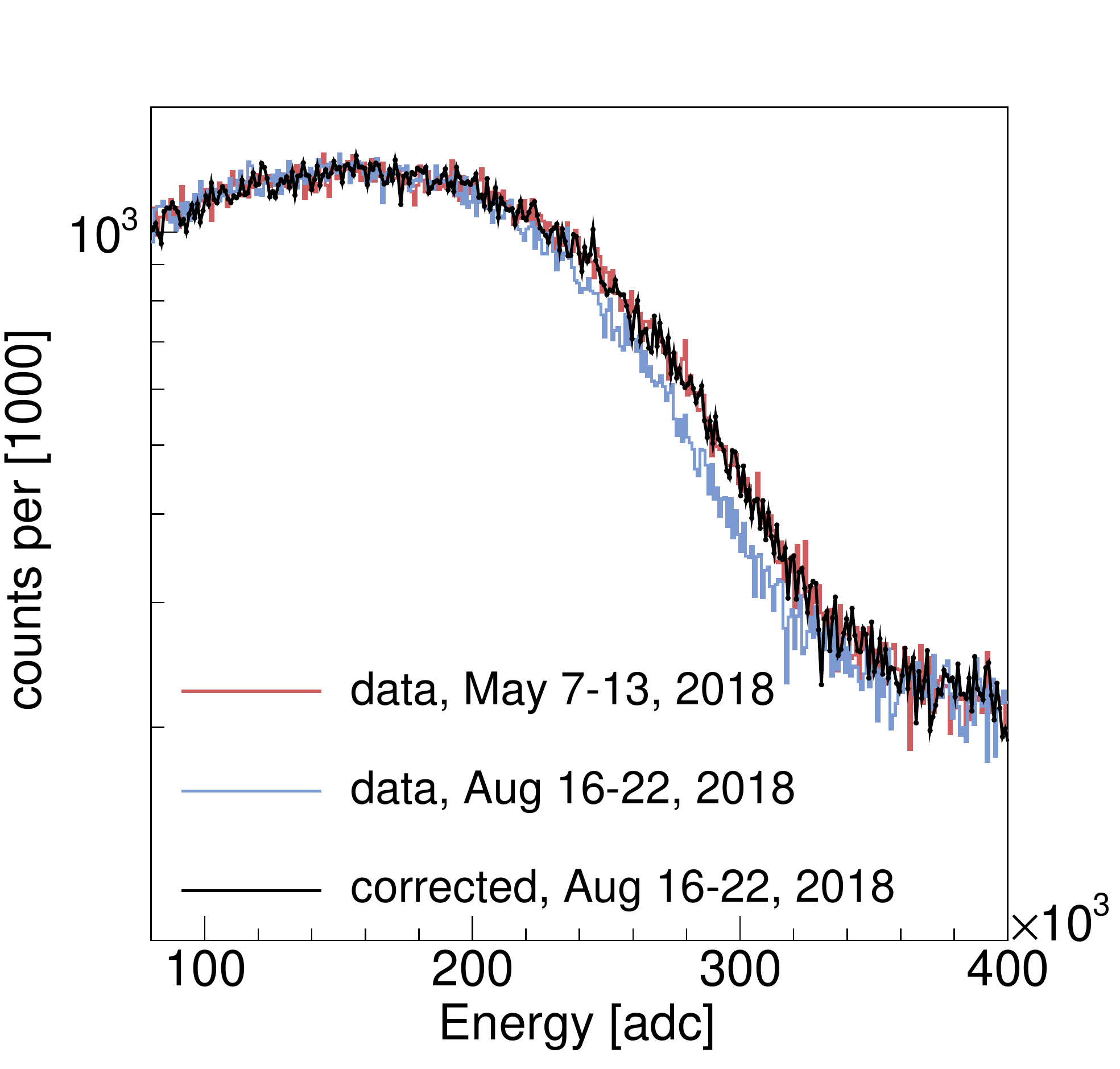}
\includegraphics[width=0.49\textwidth]{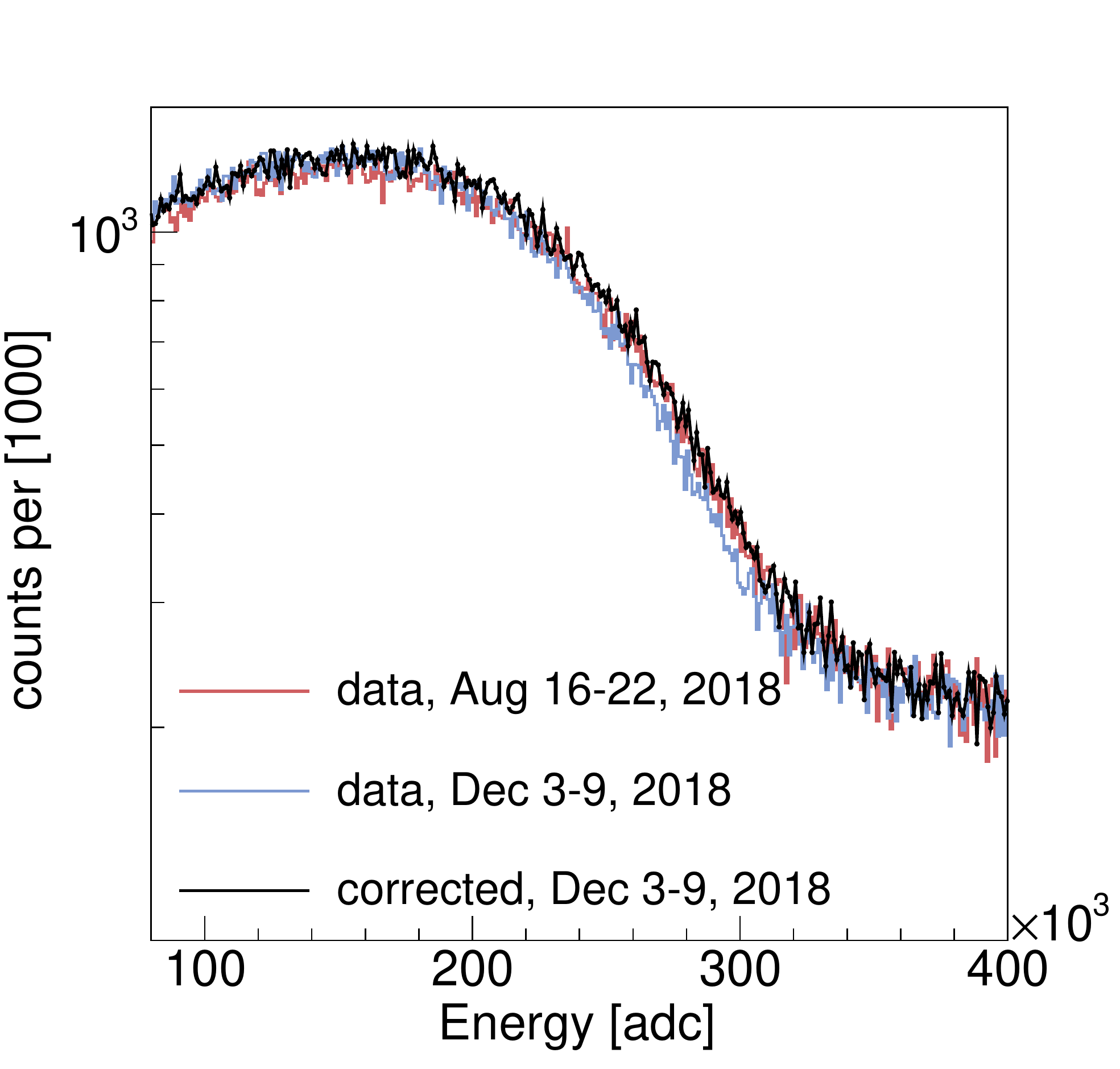}
\caption{Michel electron spectrum shift from the start (red) to the end (blue) 
of each beam-on period due to a gradual hardware gain reduction. 
The gain-corrected spectrum corresponding to the end of each beam-on 
period is presented in black.}
\label{f:gainCorrection}
\end{figure}

This prominent Michel electron peak can also be used to evaluate the magnitude 
of gain variations over the data-collection period. 
To avoid being affected by the spectral shifts due to the HOG background, 
the gain changes are evaluated using beam-off data. The left panel 
of Fig.~\ref{f:gainCorrection} shows the effective gain decrease of the August 16-22 
beam-off data with respect to the May 7-13 beam-off data. To compute the relative 
gain-correction factor $g_c$ between those two periods, we minimize the \(\chi^2\) between 
the May 7-13 spectrum and the $g_c$-scaled August 16-22 spectrum, resulting in  
\(g_c = 1.041\). Similarly, the right panel of Fig.~\ref{f:gainCorrection} 
shows the effective gain decrease of the December 3-9 beam-off data with respect 
to the August 16-22 beam-off data, for which \(g_c = 1.026\). 
However, it is not possible to correct for gain changes during beam-on periods 
due to the HOG-background effect on the spectrum, as will be shown in the next section. 
Instead, the muon-induced neutron background is used to 
derive the change in the \(E_2\) cut over time.

\subsection{\label{ss:muonN_selection}Capture energy cut selection from 
the muon-induced neutron events}

Cosmic muons also produce spallation neutrons in the detector and in surrounding 
materials like the concrete walls~\cite{Hertenberger:1995ae, Boehm:2000ru,
PhysRevD.64.013012, osti_1151758}. The detection of this neutron 
population constitutes a steady-state background that does not depend 
on the SNS beam status, and thus, can be employed to derive the \(\Delta t\) and 
\(E_2\) neutron-capture cuts for the detection of the beam neutrons. 
In this section, we describe a procedure to adjust the \(E_2\) neutron-capture cuts 
in order to compensate for time-dependent spectral shifts present in our data.  
Our method derives the time-dependent \(E2\) cuts that match the  
measured muon-induced neutron rate to a chosen reference value.

As in the case of Michel electrons, the muon-induced neutron 
interactions are selected by tagging events with 
high energy depositions from muon ionization. 
While the muon-induced neutrons would eject prompt protons 
before the Gd neutron-capture gamma-ray emission, 
triple-pulse events that would correspond to 
a muon track followed by prompt proton recoils and a Gd gamma-ray shower
are not observed. On the other hand, 
we detect a useful rate of pulse pairs with \(E_1 > 3 \times 10^{5}\)~adc
followed by \(E_2\) pulses with the \(\Delta t\) time structure consistent with 
neutron capture in Gd. These represent muon-track high-energy depositions, 
possibly piling up with prompt proton recoils from spallation neutrons 
either created in the detector or in the surrounding materials, followed by 
their capture in Gd. 

\begin{figure}
\centering
\includegraphics[width=0.49\textwidth]{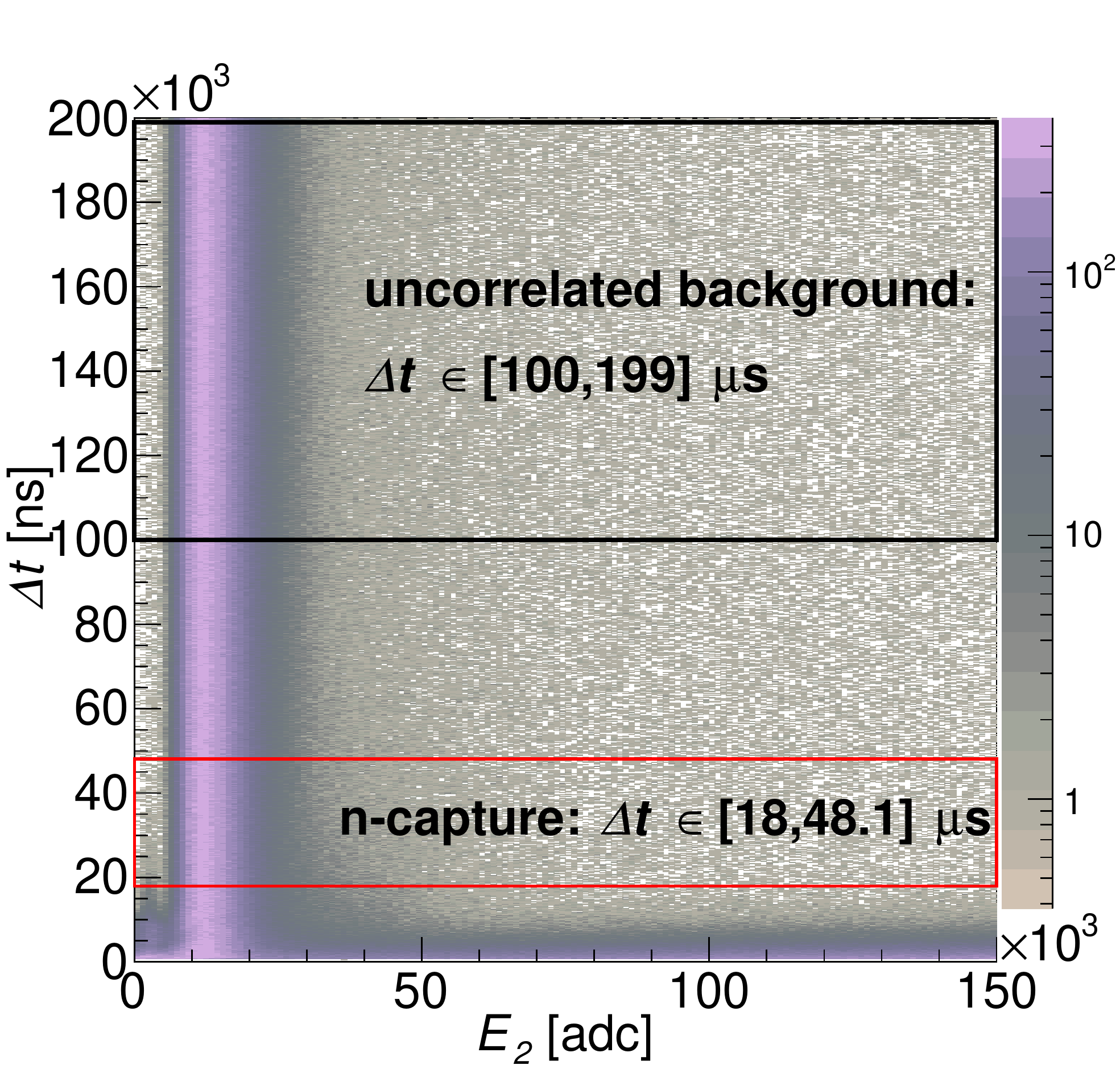}
\includegraphics[width=0.49\textwidth]{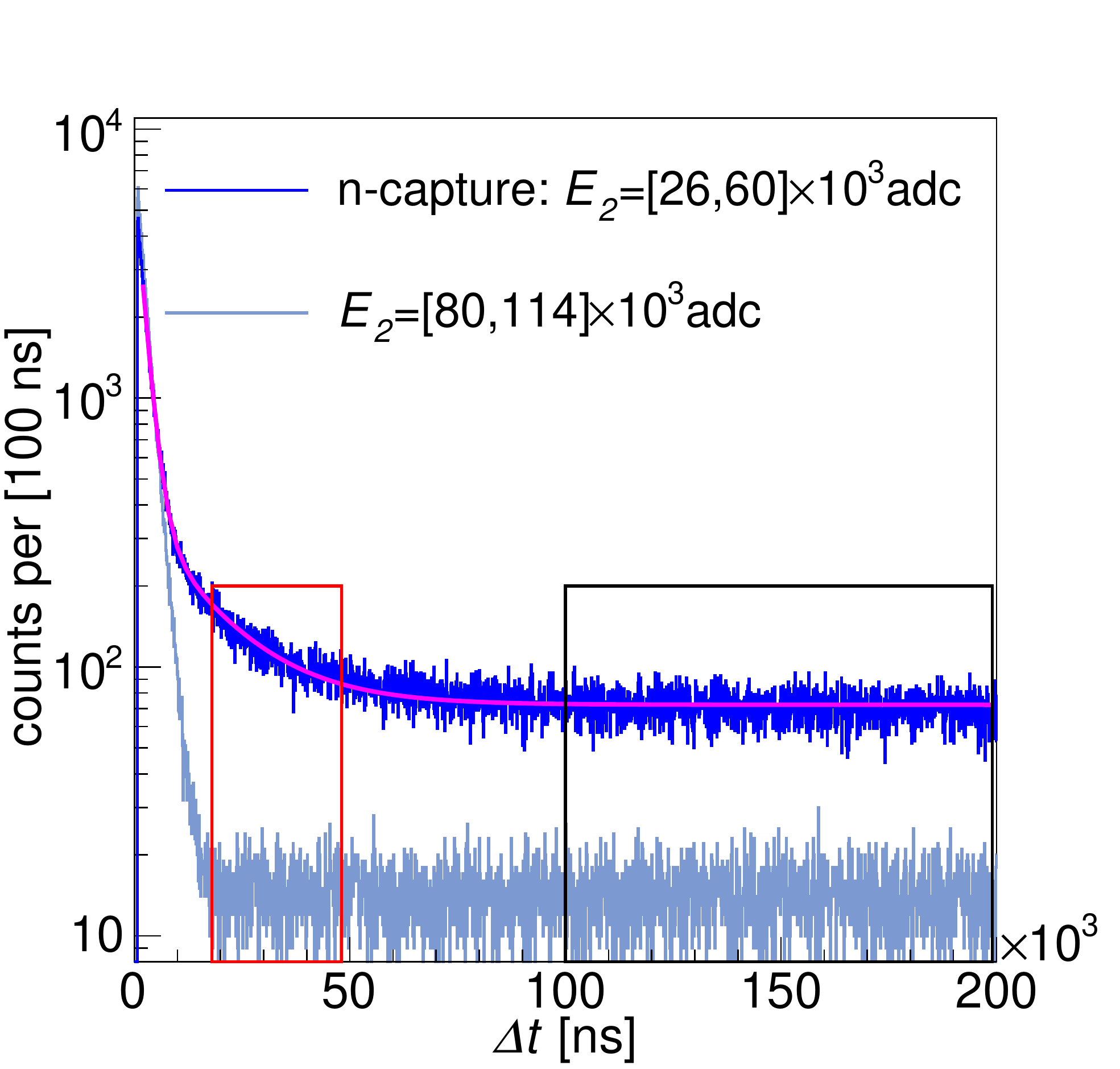}
\caption{Left: Histogram $H_T(E_2,\Delta t)$ of the inter-pulse time \(\Delta t\) versus 
the second pulse's energy \(E_2\) 
with the first pulse's energy \(E_1 > 3 \times 10^{5}\)~adc 
selected to represent muon tracks for the April 2018 beam-off data. 
Right: \(\Delta t\)-projections for two different \(E_2\) 
ranges. A double exponential fit (pink) of the \( E_2  \in [26,60] \times 10^{3}\) 
adc \(\Delta t\)-projection results in \(\tau_{\mu -} = (1.99 \pm 0.01)~\mu s\) 
and \(\tau_{Gd} = (15.49 \pm 0.41)~\mu s\). The neutron-capture ranges 
are labeled as n-capture. }
\label{f:muonSelection_dt}
\end{figure}

The left panel of Fig.~\ref{f:muonSelection_dt} contains the 
$\Delta t$-vs.-$E_2$ histogram $H_T(E_2,\Delta t)$ of pairs for which the first pulse's energy is constrained 
to \(E_1 > 3 \times 10^{5}\)~adc in order to select muon-track energy depositions. 
Since the background rate is higher at low energies, this 2D histogram
shows a high-rate band of uncorrelated events near the energy threshold. 
The right panel of Fig.~\ref{f:muonSelection_dt} shows the \(\Delta t\)-projections 
for two different \(E_2\) ranges: one range is overlapping with the region 
of $E_2$ depositions resulting from neutron captures --- as will be shown 
below --- and the other range is selected well above the neutron-capture 
$E_2$ region. While the Michel electron events are irreducibly present in both ranges, 
time-correlated events with a time constant fit value \(\tau_{Gd} = (15.49\pm 0.41)~\mu s\) are 
present when we set \( E_2  \in [26,60]\times 10^{3}\) adc, corresponding to muon-induced 
neutron captures. 

\begin{figure}
\centering
\includegraphics[width=0.49\textwidth]{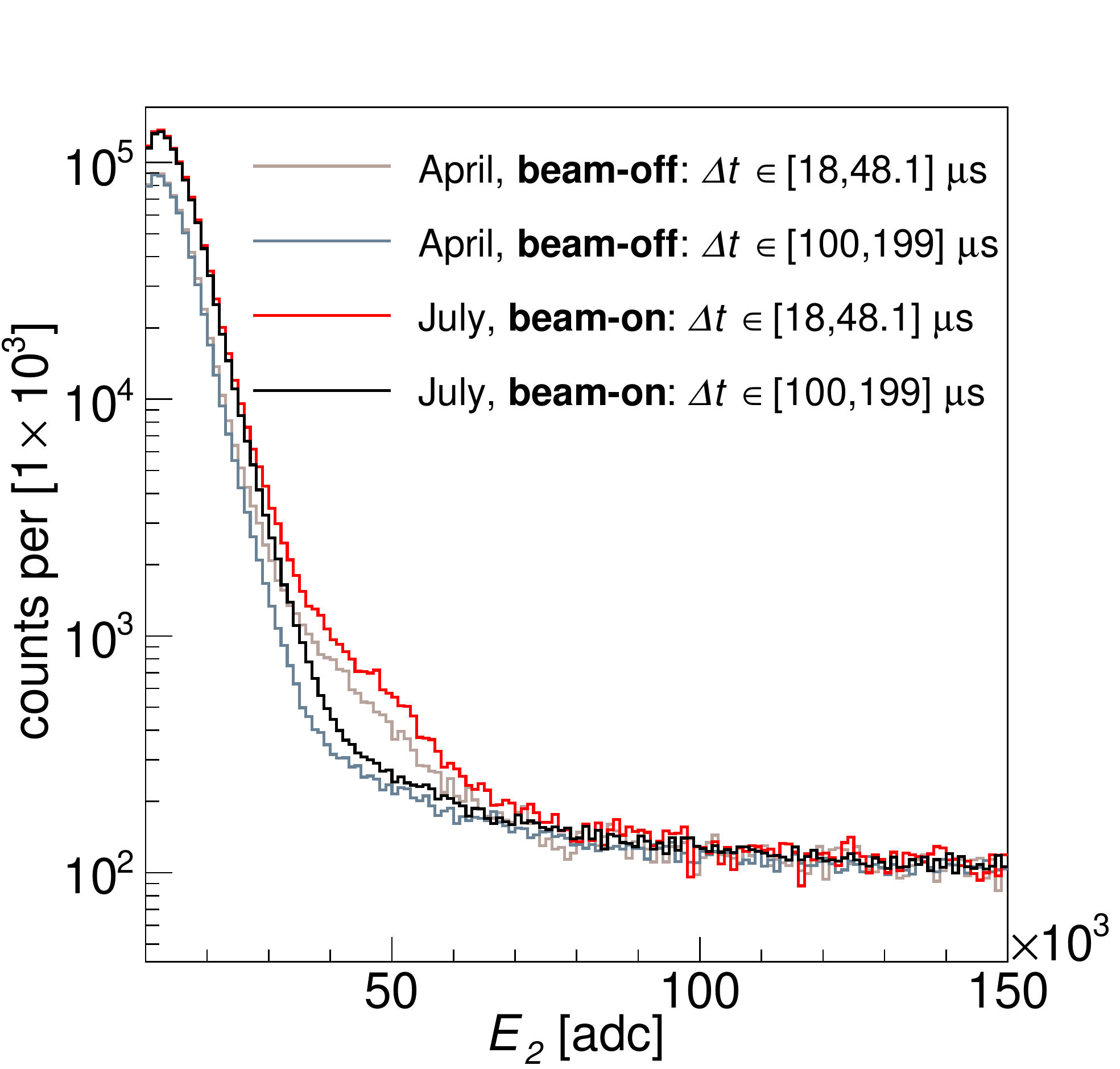}
\includegraphics[width=0.49\textwidth]{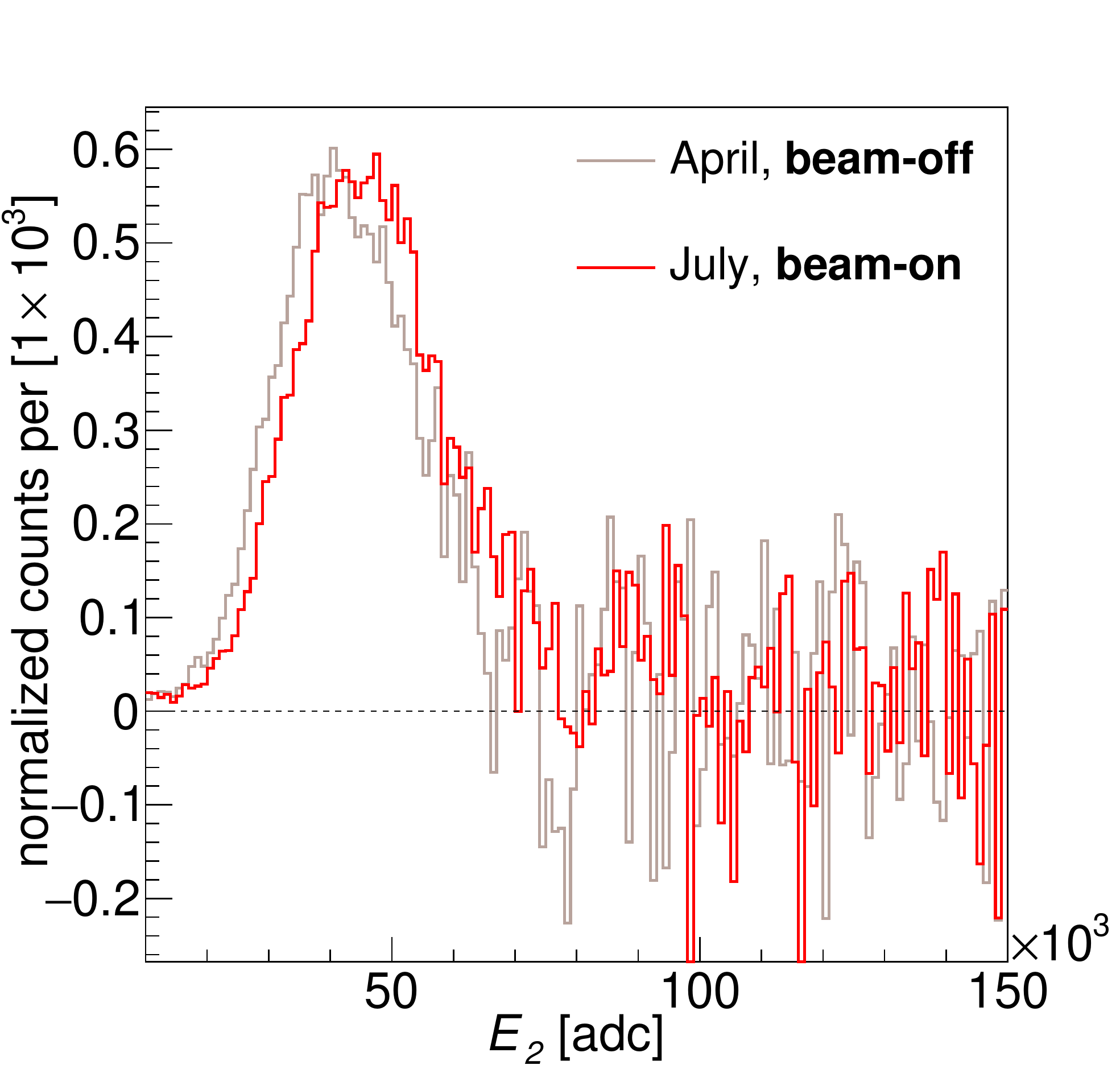}
\caption{Left: \(E_2\)-projections for the two \(\Delta t\) ranges highlighted in 
the left panel of Fig.~\ref{f:muonSelection_dt}, for the April 2018 beam-off 
and the July 2018 beam-on data. Right: residual counts between the two projections 
normalized by the neutron-capture projection bin content, for the April 2018 
beam-off and the July 2018 beam-on data.
Although the \(E_1\) values should also shift as \(E_2\) with the HOG background, 
using the same \(E_1 > 3 \times 10^{5}\)~adc cuts  for beam-on and beam-off data 
yields similar residuals, \textit{i.e.}, similar muon-induced neutron-capture counts. }
\label{f:muonSelection_E2}
\end{figure}

The left panel of Fig.~\ref{f:muonSelection_E2} shows the \(E_2\)-projections 
for the two different \(\Delta t\) ranges marked in the 2D histogram of Fig.~\ref{f:muonSelection_dt}.  
The neutron-capture \(\Delta t\) range, enclosed in a red bordered rectangle 
in Fig.~\ref{f:muonSelection_dt}, is taken 
to start at \(\sim 8 \times \tau_{\mu -}\) in order to reject Michel electron pairs, 
and ends at \(\sim 3 \times \tau_{Gd}\). The uncorrelated 
\(\Delta t\) range, enclosed in a black bordered rectangle, represents the flat distribution of 
accidental, hence uncorrelated, background event pairs.  
The residual counts between these two $E_2$ histograms--- after properly 
scaling the background histogram to account for its larger \(\Delta t\) projection range---
constitute the muon-induced neutron-capture event count, that, as expected, has the same 
magnitude for a beam-off month (April 2019) and a beam-on month (July 2019). 
However, as illustrated in the right panel of Fig.~\ref{f:muonSelection_E2}, the \(E_2\) 
values of the beam-on residual counts are shifted up compared to the beam-off 
\(E_2\) values by \(\sim 5 \times 10^{3}\) adc, which corresponds to about 
\(0.5~\text{MeVee}\) in light-output energy units. This systematic increase 
in the pulse-integral values is due to a high pile-up rate of HOG-background 
\(511\)~keV gamma rays. 

\begin{figure}
\centering
\includegraphics[width=0.7\textwidth]{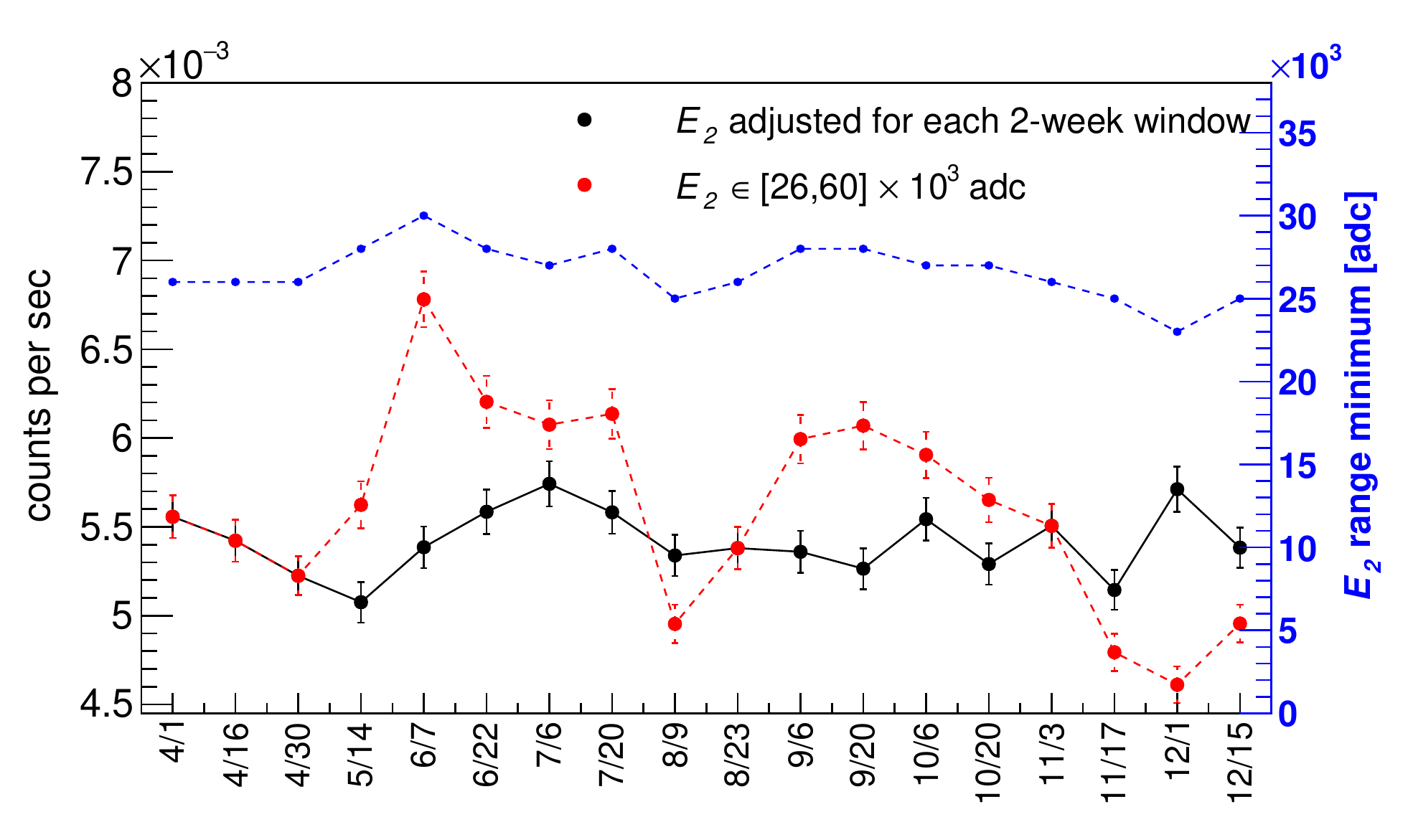}
\caption{Muon-induced neutron-capture counts integrated over two-week 
intervals for \(E_2\) cuts adjusted to maintain a \enquote{constant} event 
count, with mean and standard deviation equal to \((5.4 \pm 0.2)\times 10^{3}\).
As a comparison, the counts for a fixed \(E_2\) cut are also plotted. 
Note that the first three points correspond to two-week intervals overlapping with the  
April 1-May 6 reference period, and thus, the derived \(E_2\) cuts did not change. 
The \(E_2\) range minima adjusted for each two-week interval are plotted in 
blue.}
\label{f:muon_signal_2week_variableE2}
\end{figure}

As a consequence, the muon-induced neutron-capture counts integrated 
within a fixed \(E_2\) range vary over time depending on the intensity 
of the HOG background. That behavior is demonstrated by the red plot of 
Fig.~\ref{f:muon_signal_2week_variableE2}, where the two-week muon-induced 
neutron-capture counts, computed for the fixed cut 
\(E_2 = [26,60] \times 10^3\) adc, show a similar profile over time
as the PMT rates of Fig.~\ref{f:beampower_rates_sizes}. 

The number of muon-induced neutron events produced in MARS is independent of beam 
power, though it would depend on changes in the overburden and in the cosmic-ray 
rate. Data from ground-based detectors that monitor cosmic-ray activity~\cite{Clem2000} show 
less than 1\% variation in the year 2018~\cite{NMDB}. Thus, since MARS remained under the 
same overburden during the 2018 data run, we can employ muon-induced neutron 
events to adjust the \(E_2\) neutron-capture cuts for each two-week data interval.

To determine the reference muon-induced neutron rate, we select the average 
of five beam-off weeks from April 1 to May 6. The reference \(E_2\) and \(\Delta t\)
cuts are derived by maximizing the signal-to-noise ratio 
\(S_0 =(T-B)/\sqrt{\sigma_{T}^2+\sigma_{B}^2}\). 
In this expression, $T$ is the total pair count within the to-be-determined \(E_2\) and \(\Delta t\)
neutron-capture reference cuts, denoted as \(\mathcal{E}_\text{ref}\) and 
\(\mathcal{T}_{\text{ref}}\) respectively.
$B$ is the corresponding uncorrelated background count estimation determined for 
the same \(\mathcal{E}_\text{ref}\) range but using \(\Delta t = [100,199]~\mu s\), 
scaled to account for the difference in \(\Delta t\) range size. 
Imposing the constraint \(\Delta t > 18~\mu s\) to preclude 
muon-Michel electron pairs from affecting the cuts, 
the signal-to-noise ratio $S_0$ is maximized for 
\(\mathcal{E}_\text{ref}= [26,60] \times 10^3\)~adc and \(\mathcal{T}_{\text{ref}} = [18,48.1]~\mu s\).
Using the rough energy scaling discussed above, the 
\(\mathcal{E}_{\text{ref}}\)  range falls around  
\([2.6, 6]\)~MeVee, which overlaps with the Gd gamma-shower range
after folding in the detector response. 

\begin{figure}
\centering
\includegraphics[width=0.49\textwidth]{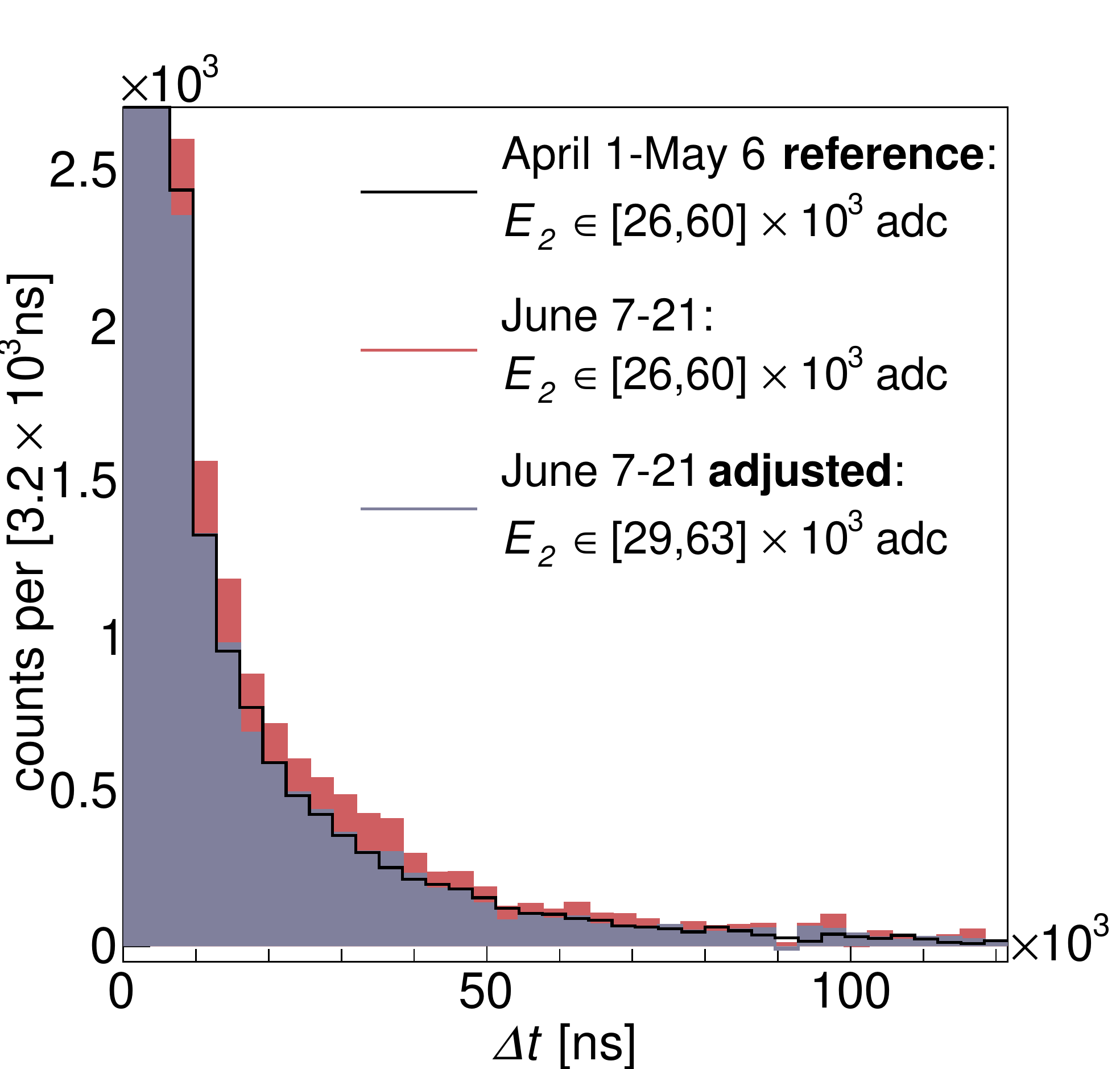}
\caption{Background-subtracted \(\Delta t\)-projections of pairs with the prompt 
pulse \(E_1 > 3 \times 10^{5}\)~adc selected to represent muon tracks,
from April 1-May 6 beam-off and June 7-21 beam-on data. The June 7-21 
\(\Delta t\)-projection for the adjusted \(E_2\) cuts matches the April 1-May 
6 reference \(\Delta t\)-projection.}
\label{f:muon_dtsignal_E2Calibration}
\end{figure}

With the $\Delta t$ cut set to $\mathcal{T}_{\text{ref}}$, the cut on $E_2$ is derived for each 
two-week interval so that the same muon-induced neutron rate as in the reference 
period is maintained. To do this, we first create the background-subtracted 
$\Delta t$-vs.-$E_2$ histogram $H_S(E_2,\Delta t)$. The mean uncorrelated 
background $B(E_2)$ per $\Delta t$ bin as a function of the $E_2$ bins is computed 
from the \(\Delta t = [100,199]~\mu s\) range of the total histogram  
$H_T(E_2,\Delta t)$ of Fig.~\ref{f:muonSelection_dt}. Thus,  
$H_S(E_2,\Delta t)$ is obtained by subtracting  $B(E_2)$ from each $(E_2,\Delta t)$ 
bin of $H_T(E_2,\Delta t)$. 
Fig.~\ref{f:muon_dtsignal_E2Calibration} shows the \(\Delta t\)-projections
of $H_S(E_2,\Delta t)$ for \(E_2 \in \mathcal{E}_\text{ref}\) 
corresponding to the reference beam-off data and to two weeks of beam-on 
data. The beam-on data $\Delta t$-projection clearly shows more counts in the 
integration range \(\mathcal{T}_{\text{ref}}\). 
Next, we derive the adjusted cut \(\mathcal{E}_{\text{adj}} \equiv [E_2^{\text{min}},E_2^{\text{max}}]\) 
by minimizing the 
\(\chi^2\) between the reference and the two-week \(\Delta t\)-projection histograms, 
summing over the \(\Delta t \in \mathcal{T}_{\text{ref}}\), 
while keeping a fixed $E_2$
range width \(E_2^{\text{max}} - E_2^{\text{min}} = (60-26) \times 10^{3}\) 
adc. For the two-week beam-on data of Fig.~\ref{f:muon_dtsignal_E2Calibration}, 
adjusting the $E_2$ range to \([29,63] \times 10^3\)~adc matches the 
muon-induced neutron rate to the reference value. 

The two-week net counts after applying the $\mathcal{E}_{\text{adj}}$ cuts 
computed for each two-week data interval are shown in
Fig.~\ref{f:muon_signal_2week_variableE2}. The result 
demonstrates that the calculated $\mathcal{E}_{\text{adj}}$ cuts appropriately offset the 
spectrum shifting effect due to the HOG background pile-up and therefore 
can be used to compute the beam-induced neutron rates. Moreover, these $\mathcal{E}_{\text{adj}}$ cuts 
also correct for the gradual gain drift that appears to have 
occurred over the several months of 2018 MARS operation; they would also 
correct for other possible PMT-response changes that may have occurred during 
the beam-on periods but that would be masked by the variability of the 
HOG-background shifting effect on the spectrum. The $E_2^{\text{min}}$ 
computed for each two-week interval, also shown in Fig.~\ref{f:muon_signal_2week_variableE2},
is distributed with a relative standard deviation 
\(\sigma_{E_2^{\text{min}}}/\mu_{E_2^{\text{min}}} = 6\)~\%, 
which represents an estimate of the additional energy smearing due to 
gain variations and variability in the HOG-background-induced spectral shifts 
when all the 2018 beam-on data is considered. An advantage of this approach
is that, as long as the detector data have been collected under the same 
overburden and incident cosmic-ray rate of the reference period, 
the same muon-induced neutron reference rate can be used to determine the 
 $E_2$ cuts in adc units, since such adjusted ranges 
will correspond to equivalent physical energy ranges in MeVee units. 

\section{\label{s:analysis}Measured BRN rate} 

\begin{figure}
\centering
\includegraphics[width=0.49\textwidth]{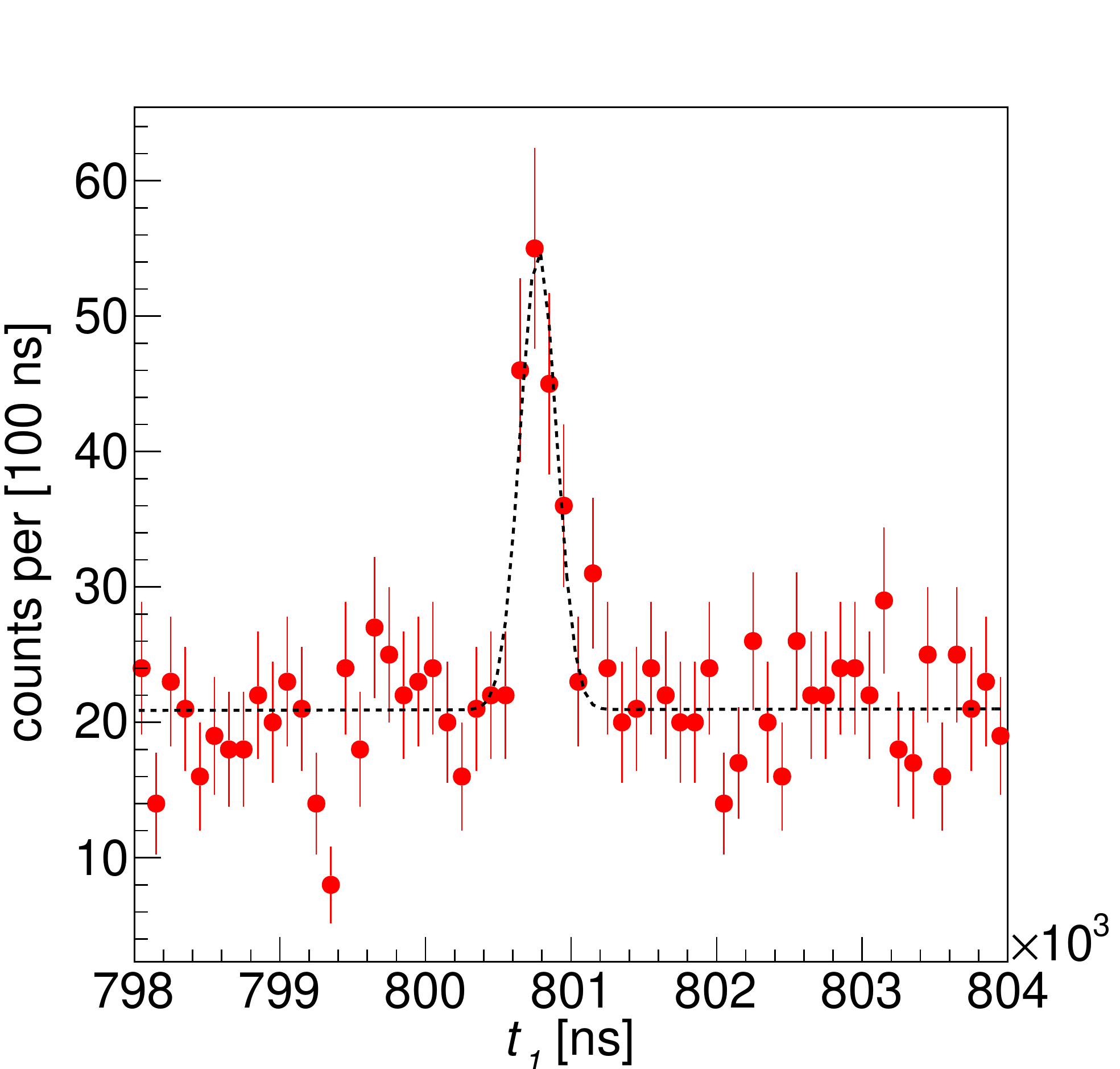}
\caption{
Histogram of the time interval \(t_1\) between the first pulse of the pair, 
or prompt pulse, and an SNS-provided 60-Hz signal, named event-39.  
The histogram entries correspond to event pairs from the May-December 2018 data 
satisfying the neutron-capture cuts for $\Delta t$ and $E_2$  
($\Delta t \in \mathcal{T}$ and $E_2$ obeying the two-week $\mathcal{E}_{\text{adj}}$ cuts), 
and with a preceding 
SNS-provided signal, named event-61, supplied during 
beam-on operation whenever there is a POT event.
The x-axis has been zoomed around the time window \( T_{\text{beam}} = [800.0, 802.0] ~\mu s\) 
that contains the beam-neutron candidate pairs. 
The Gaussian fit results in a full width at half maximum 
FWHM = \(298 \pm 50\)~ns, consistent with the SNS beam time profile. 
}
\label{f:t1_dtCapt_E2Capt_ev61_MayDec2018}
\end{figure}

Any SNS-target neutrons produced during beam POT events and 
that reach MARS would interact within a \( 2~\mu\text{s}\) \enquote{beam} time window 
\( T_{\text{beam}} \), preceded by an event-61 signal. 
The event-39 timestamp is offset by \( 800~\mu\text{s}\) in the acquisition software 
so that the beam-neutron 
candidate pairs are constrained by \( t_1 \in{T_{\text{beam}}} = [800.0, 802.0] ~\mu s\). 
The beam-on data \(t_1\) histograms of Fig.~\ref{f:t1_dtCapt_E2Capt_ev61_MayDec2018} 
clearly shows the beam neutron spike at \( T_{\text{beam}} \) for pairs 
with \(\Delta t\) and \(E_2\) within cuts consistent with Gd neutron capture. 
In order to estimate the steady-state background events in \( T_{\text{beam}} \), 
we compute the number of pairs in a long out-of-beam strobe window 
\( t_1 \in{T_{\text{strobe}}}= [1,15]~ms\), which is justified by the 
flat count profile within that window. 
A third \( 2~\mu\text{s}\) out-of-beam window 
\( T_{\text{check}} = [1000.0, 1002.0] ~\mu s\) illustrates 
 that steady-state background statistical fluctuations 
are consistent with the mean background from the \( T_{\text{strobe}} \)
and do not explain the \( T_{\text{beam}} \) neutron pair excess obtained 
below. 

\begin{figure}
\centering
\includegraphics[width=0.8\textwidth]{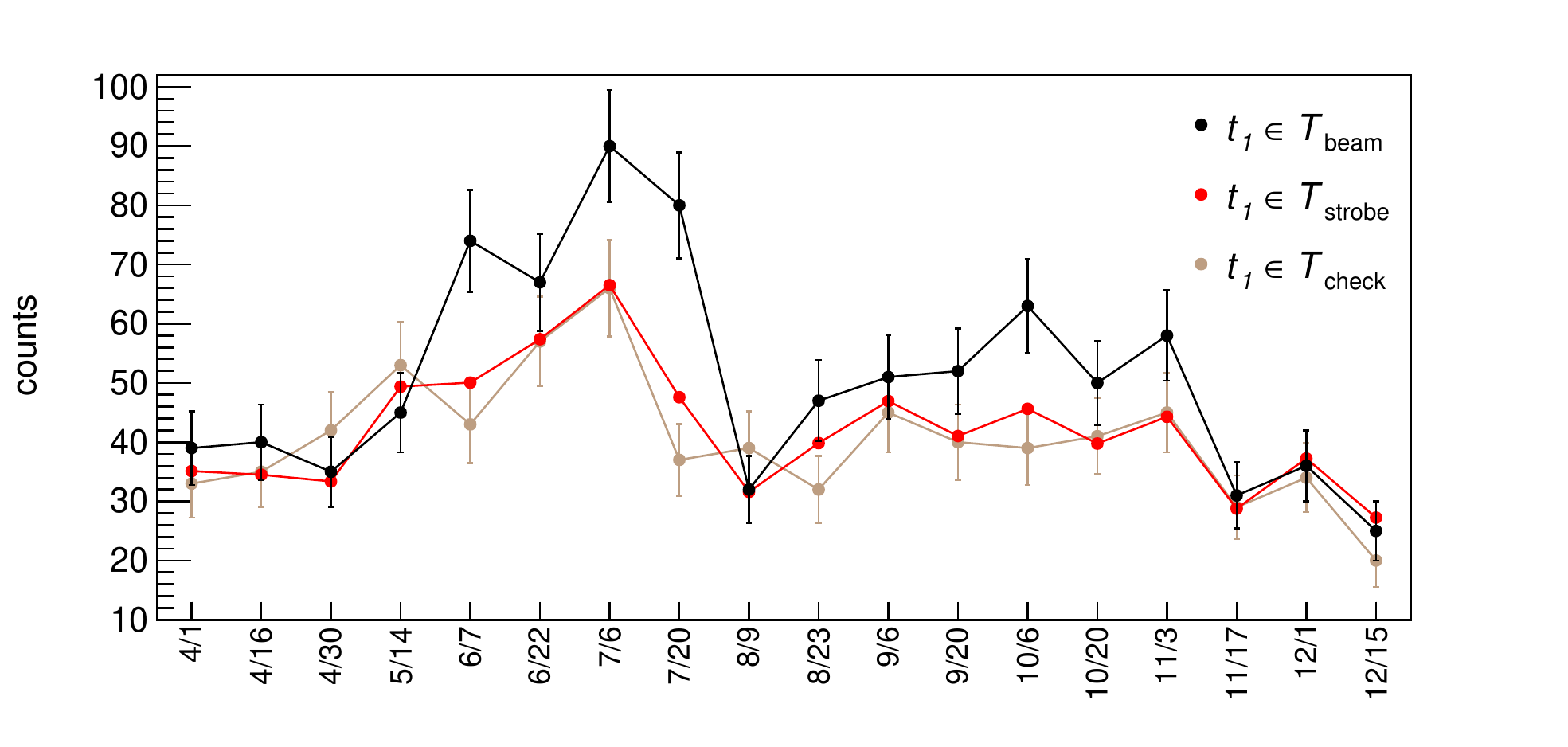}
\caption{The two-week interval counts for event pairs with 
\(\Delta t \in \mathcal{T}\), satisfying the two-week $\mathcal{E}_{\text{adj}}$  
cuts and \(t_1\) restricted to time windows \(T_{\text{beam}}\), 
\(T_{\text{strobe}}\) and \(T_{\text{check}}\). The \(T_{\text{strobe}}\) 
counts have been scaled by \(T_{\text{beam}}/T_{\text{strobe}}\). Error bars are statistical.}
\label{f:beamCounts_variableE2}
\end{figure}

Fig.~\ref{f:beamCounts_variableE2} shows the two-week interval counts 
for \(t_1\) restricted to time windows \(T_{\text{beam}}\), \(T_{\text{strobe}}\) 
and \(T_{\text{check}}\) respectively, \(E_2\) obeying the two-week  $\mathcal{E}_{\text{adj}}$  
cuts derived in section~\ref{ss:muonN_selection}, 
and \(\Delta t\) within the Gd neutron-capture cut \(\mathcal{T} = [6,48.1]~\mu s\). 
The \(\Delta t\) cut was extended down to \(6~\mu s\), which is about the time 
fast neutrons take to thermalize in MARS and their Gd capture 
probability to reach its highest value~\cite{BOWDEN2012209, roecker_2016}. 
To estimate the steady-state background 
contribution during the beam window, the \(T_{\text{strobe}}\) counts 
have been scaled by \(T_{\text{beam}}/T_{\text{strobe}}\). For both beam 
operating periods (May 14-August 5 and August 23-November 17), the total 
counts in \(T_{\text{beam}}\) are consistently larger than the average 
background counts estimated from \(T_{\text{strobe}}\). The counts in 
\(T_{\text{check}}\) are, on the other hand, consistent with background 
fluctuations.  

\begin{figure}[!htb]
\centering
\includegraphics[width=0.7\textwidth]{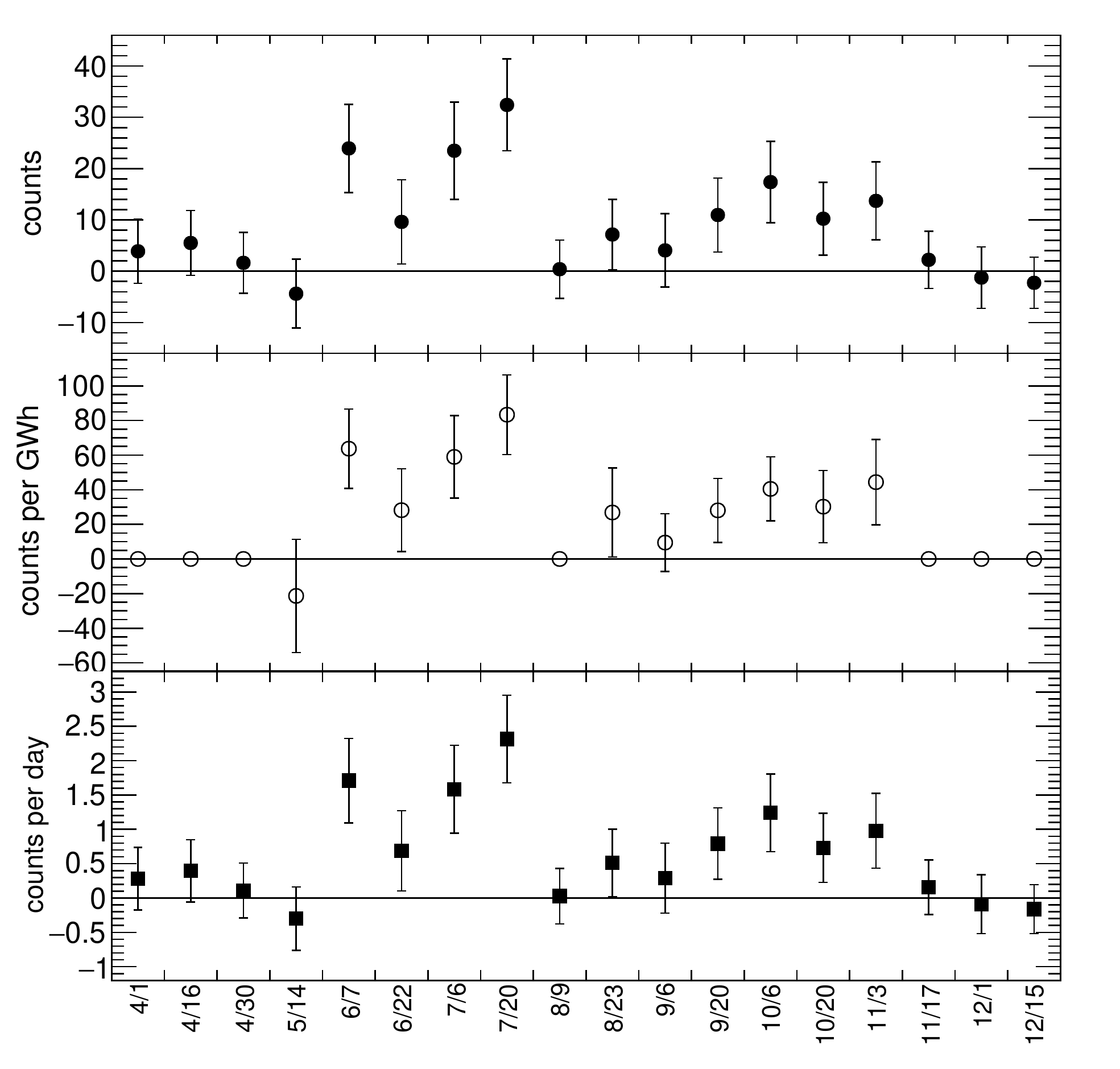}
\caption{Top: the two-week-interval residual between the \(T_{\text{beam}}\) counts 
and the scaled \(T_{\text{strobe}}\) counts. Middle: residual counts normalized by 
the integrated beam power delivered during each corresponding two-week data-integration 
interval. Bottom: residual counts normalized by the exact duration of each data integration 
interval. Error bars are statistical.}
\label{f:beamSignal_variableE2}
\end{figure}

The residual between the \(T_{\text{beam}}\) counts and the scaled 
\(T_{\text{strobe}}\) counts, representing our estimate of the beam neutron 
counts in MARS, is presented in the top panel of 
Fig.~\ref{f:beamSignal_variableE2}. Since the beam neutron rate is 
expected to be correlated with beam power, the counts in the middle panel
 of Fig.~\ref{f:beamSignal_variableE2} have been normalized by 
the integrated beam power delivered during each corresponding two-week 
data-integration interval; if the delivered energy is zero, the beam-energy-normalized 
counts are set to zero. The bottom panel of Fig.~\ref{f:beamSignal_variableE2} 
presents the counts normalized instead by the duration of 
each data-integration interval.



\begin{figure}
\centering
\includegraphics[width=0.49\textwidth]{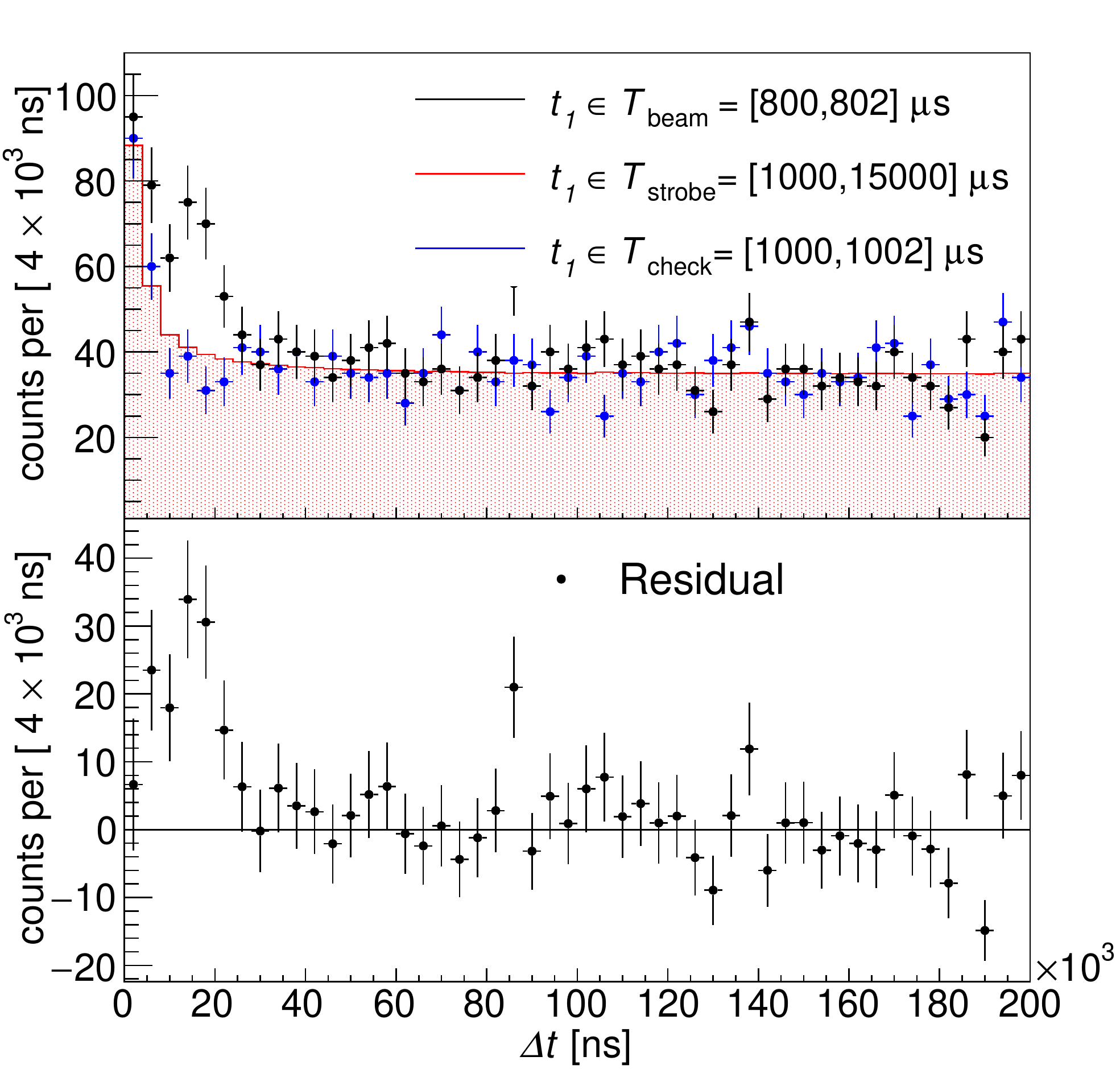}
\includegraphics[width=0.49\textwidth]{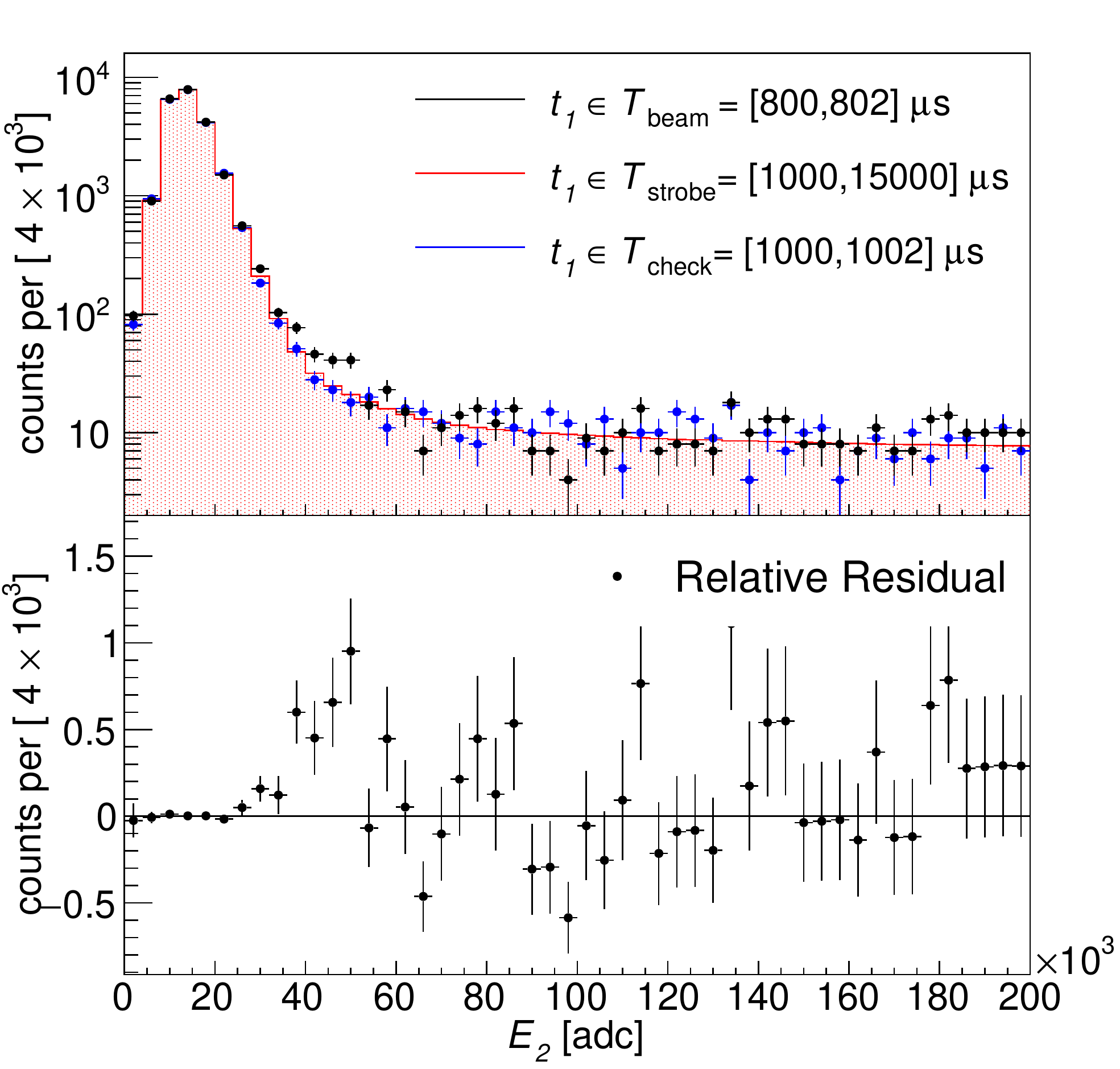}
\caption{Top left: the \(\Delta t\) histogram for event pairs satisfying the two-week 
$\mathcal{E}_{\text{adj}}$  
cuts and with a preceding event-61 signal. 
Top right: the \(E_2\) histogram for event pairs with $\Delta t \in \mathcal{T}$
and with a preceding event-61 signal. 
Bottom left: residual between the  \(T_{\text{beam}}\) and the  \(T_{\text{strobe}}\) 
\(\Delta t\)-histograms. 
Bottom right: residual between the  \(T_{\text{beam}}\) and the  \(T_{\text{strobe}}\) 
\(E_2\)-histograms, divided by the \(T_{\text{beam}}\) histogram bin content. 
All the May-December 2018 data is included.}
\label{f:dt_E2Capt_ev61_MayDec2018}
\end{figure}

Fig.~\ref{f:dt_E2Capt_ev61_MayDec2018} 
presents the \(\Delta t\) and \(E_2\) histograms for all the 2018 data, after 
respectively applying the neutron-capture 
 $\mathcal{E}_{\text{adj}}$  and \(\mathcal{T}\) cuts, and for the three \(t_1\) windows 
under consideration. The \enquote{\(T_{\text{beam}}\)} histograms 
consistently show event excesses within the respective neutron-capture 
regions, compared to the background histograms for the 
\(T_{\text{strobe}}\) and \(T_{\text{check}}\) windows. The \(\Delta t\) 
residual excess, shown in the bottom panel of Fig.~\ref{f:dt_E2Capt_ev61_MayDec2018},
 seems to extend only up to \(\sim 30\)~\(\mu\)s, which is 
shorter than the \(\mathcal{T}\) cut, indicating that this cut 
can be further tuned to potentially enhance the signal-to-background ratio. 
The \(E_2\) residual excess, on the other hand, seems to end somewhere 
around \(6 \times 10^4 \)~adc, closely matching the adjusted 
\(\mathcal{E}\) ranges.

\begin{figure}
\centering
\includegraphics[width=0.49\textwidth]{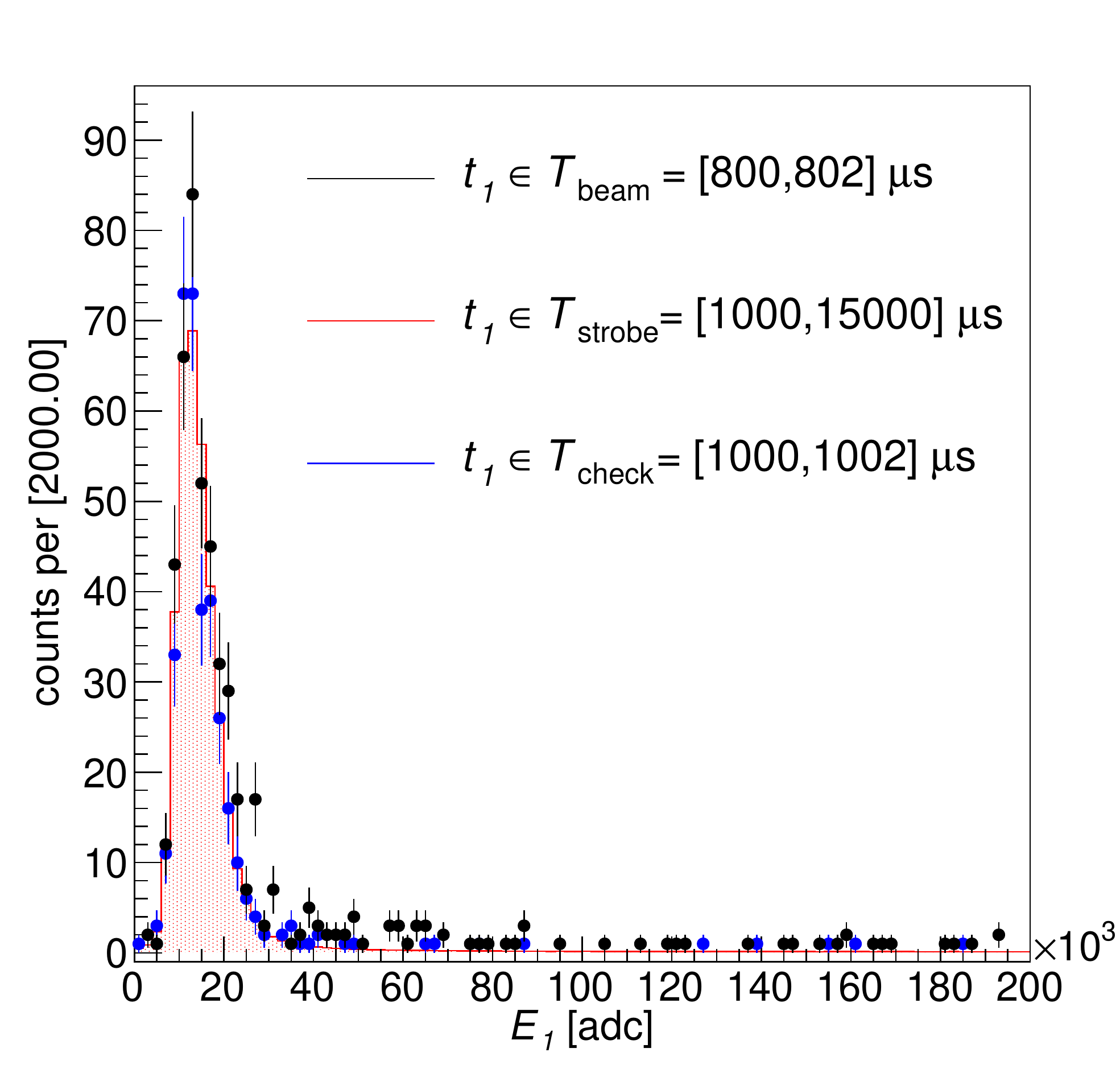}
\includegraphics[width=0.49\textwidth]{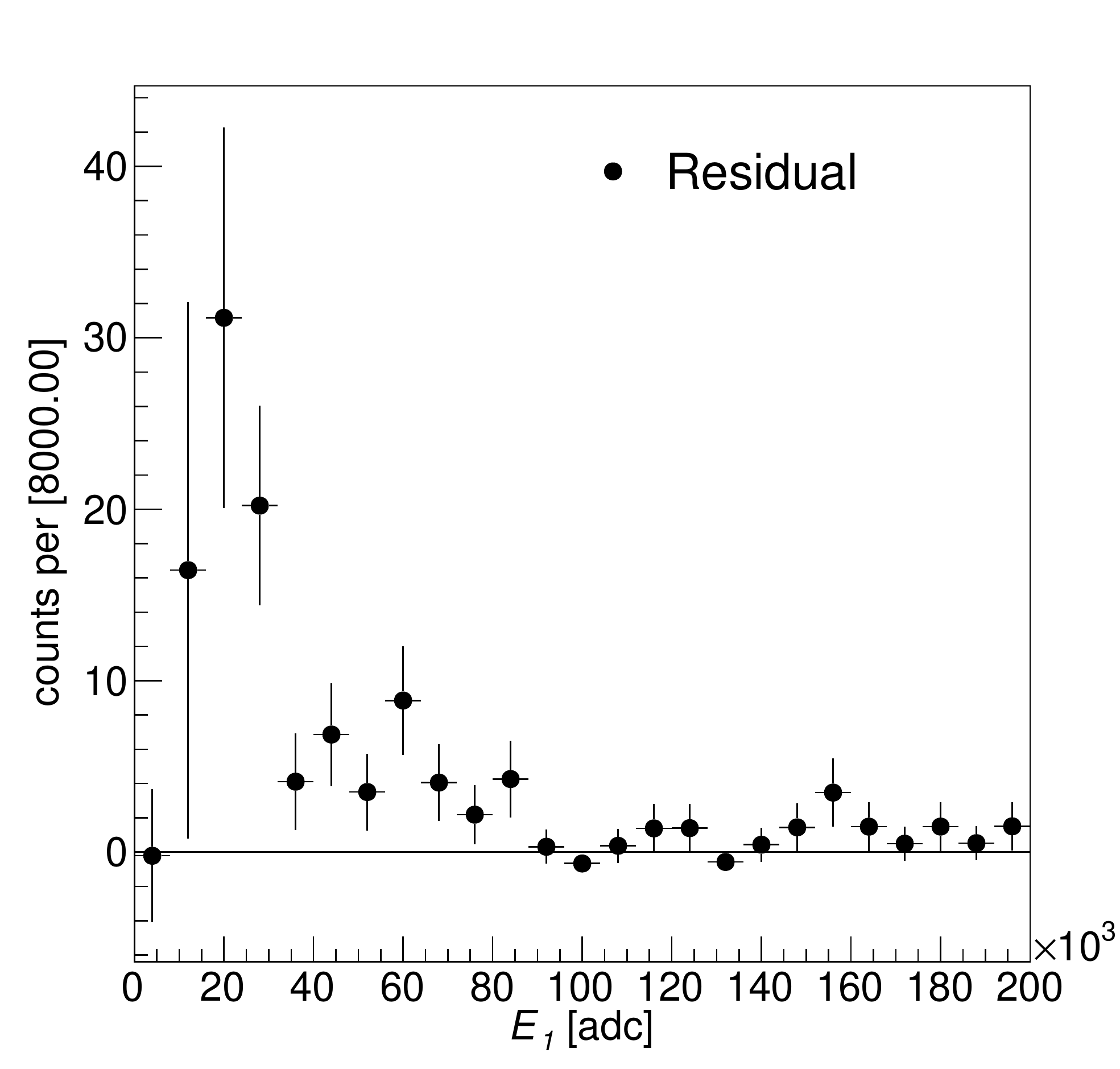}
\caption{Left: The \(E_1\) histogram for event pairs with $\Delta t \in \mathcal{T}$, 
satisfying the two-week $\mathcal{E}_{\text{adj}}$  
cuts, and with a preceding event-61 signal, 
from May-December 2018 data. Right: residual between the  \(T_{\text{beam}}\) 
and the  \(T_{\text{strobe}}\) histograms.}
\label{f:E1_dtCapt_E2Capt_ev61_MayDec2018}
\end{figure}

Lastly, the left panel of Fig.~\ref{f:E1_dtCapt_E2Capt_ev61_MayDec2018} shows the 
deposited prompt energy \(E_1\) for all the 2018 data satisfying the Gd neutron 
capture cuts. The \(E_1\) residual excess during \(T_{\text{beam}}\), 
presented in the right panel of Fig.~\ref{f:E1_dtCapt_E2Capt_ev61_MayDec2018}, 
is highly suppressed beyond \(\sim 8 \times 10^4 \)~adc, which approximately 
corresponds to \(8\)~MeVee, or $\sim 15$~MeV in proton recoil energy. 
As noted in section~\ref{ss:muonN_selection},
the gain variations and the HOG-background-induced spectral shift variability 
encountered in the 2018 beam-on data should contribute an additional \(\sim 6\)\% 
to the energy smearing of the $E_2$ and $E_1$ histograms of 
Fig.~\ref{f:dt_E2Capt_ev61_MayDec2018} and \ref{f:E1_dtCapt_E2Capt_ev61_MayDec2018}.


\section{\label{s:alleyBRNflux}BRN flux at the Neutrino Alley}



A total of 148.6 \(\pm\) 26.0 neutron counts were detected within 
the $2~\mu s$ beam window in 154 beam-on days when the total POT 
delivered energy was 
\(3.872\)~gigawatt-hour (GWh). To estimate the BRN flux incident 
on MARS from these data, we use the neutron detection efficiency
measured with a deuterium-tritium (DT) neutron generator.
Next, we describe this measurement and derive the detection 
efficiency for 14~MeV neutrons, albeit with a large uncertainty. 
A careful analysis of the DT generator data combined with
the detector-response modeling will be treated in a separate publication. 


\subsection{\label{ss:efficiency} Flux estimation from MARS data}

The MARS detector was irradiated with a Thermo API120 neutron generator 
~\cite{API120}, on October 24\textsuperscript{th} 2019 during a beam-off week. 
The generator employs the associated particle imaging (API) technique
~\cite{CHICHESTER2005753, Beyerle1990}, where the
3.5 MeV alpha particle produced by the D-T reaction in time coincidence 
and angular correlation with a 14 MeV neutron is detected by a 100 micron
thick YAP:Ce crystal mounted on a fiber-optic faceplate. The scintillation light 
is collected by a Hamamatsu H13700 multi-anode flat panel PMT coupled to a 
resistor network producing four scaled outputs enabling reconstructing of 
the position via Anger-logic~\cite{osti_4036016}. The 
direction of the associated neutron is determined within any 
one of 32 x 16 angular pixels, each subtending a 4.5 degree opening-angle cone. 

\begin{figure}
\centering
\includegraphics[width=0.49\textwidth,align=c]{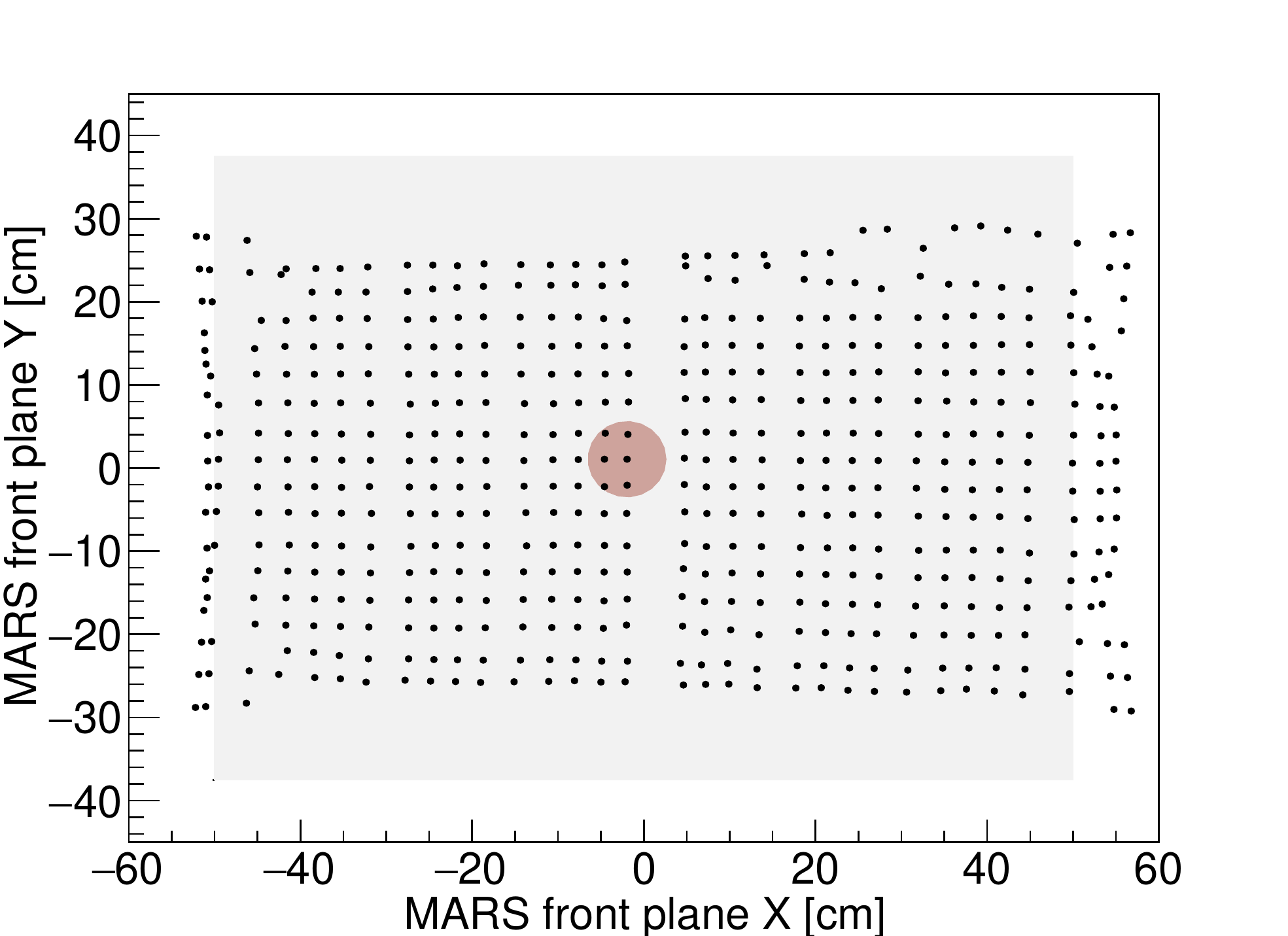}
\includegraphics[width=0.49\textwidth,align=c]{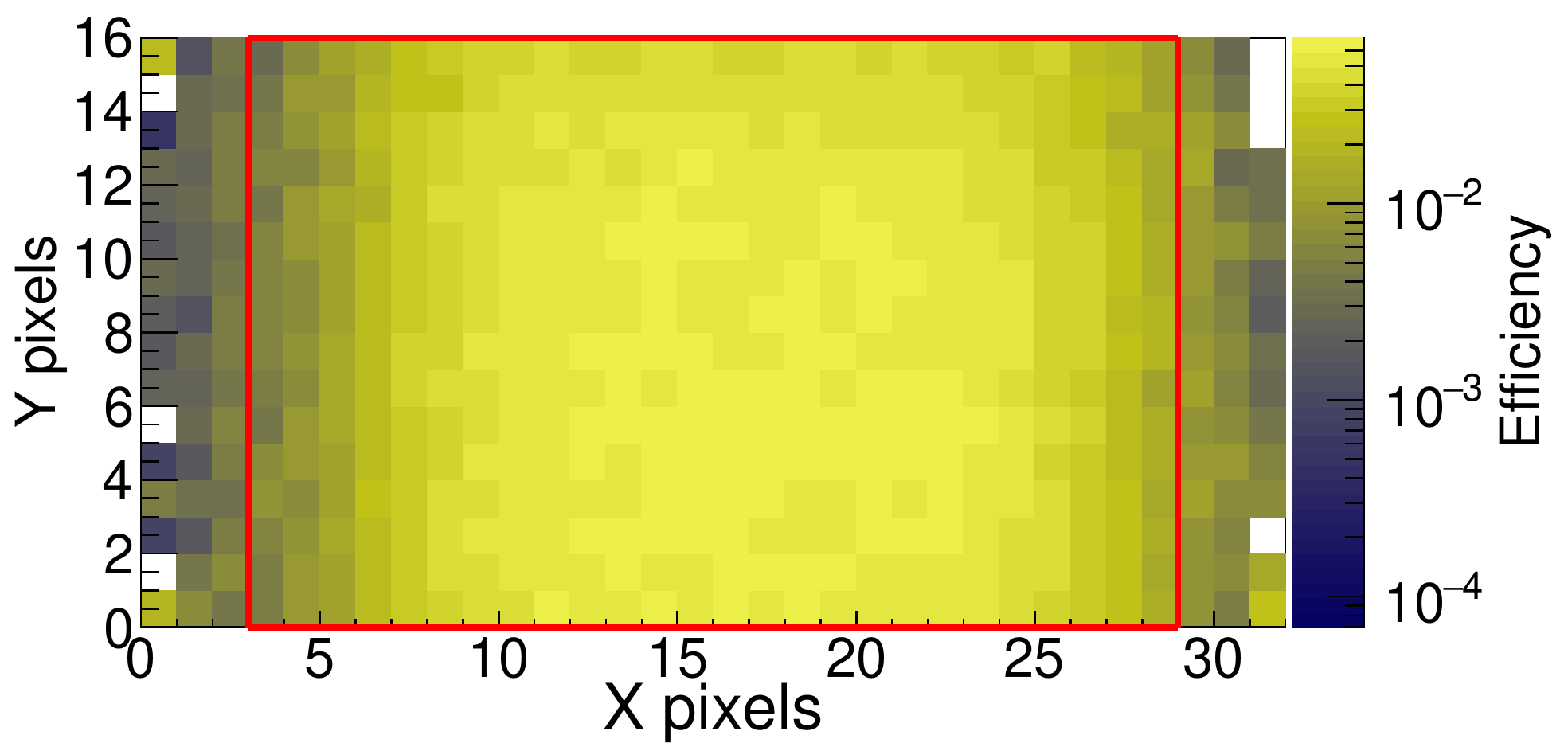}
\caption{Left: Black dots are the projection of the alpha-sensor pixel 
centers onto MARS' front plane in gray; the red circle is the projection of the 4.5 degree opening-angle 
neutron cone corresponding to the central pixel. 
Right: Efficiency computed for each alpha sensor pixel;  
the red rectangle delineates the pixels used to estimate the overall efficiency.}
\label{f:alphaPixels}
\end{figure}

The neutron-generator target was placed 117~cm perpendicular from the MARS side face, 
at three different vertical and horizontal offsets from MARS' center. 
Here, we only employ the data collected when the target was aligned with MARS' center 
such that the solid-angle region spanned by the alpha-tagged 14~MeV neutrons was within 
the horizontal edges of the MARS scintillator volume but slightly extended 
beyond the vertical edges. This is illustrated in Fig~\ref{f:alphaPixels} (left) 
which shows the projection of the alpha-sensor pixel centers onto MARS' front plane
as well as the projection of the neutron cone for only the central pixel. 
Only alpha triggers for which the collected energy is within \(\pm 20\)\% of 
the full 3.5 MeV energy are included, which rejects incorrect triggers due to 
target-emitted x-rays. 

The generator was run at its lowest power setting yielding \(\sim 10^{6}\)~n/s. 
Due to the higher event rate (\( 25\)~kHz) compared to regular operation, 
the inter-pulse window to form MARS pairs was reduced to \(\Delta t < 100~\mu s\). 
The prompt pulses generated by alpha-coincident 14~MeV neutrons create a 
sharp spike at \(t_1 \in T_{\text{DT}}=[430,490]~\text{ns}\), where 
\(t_1\) is here defined as the time from the first pulse of the pair to the preceding 
alpha trigger. 
The \(E_1\) energy distribution of fully thermalized 14~MeV neutrons peaks at 
around 4~MeVee, above the $\sim$ 1~MeVee MARS threshold and is thus not  
affected by the gain decrease observed in these data. 
The adjusted range  $\mathcal{E}_{\text{DT}} = [23,57] \times 10^{3}~\text{adc}$ 
is determined by matching 
the same reference muon-induced neutron rate of section~\ref{ss:muonN_selection}
with the muon-induced neutron rate of the beam-off week immediately before this measurement. 
The rate of uncorrelated background pairs within the signal range $\Delta t \in \mathcal{T}$ 
is determined from the uncorrelated-background range \(\Delta t \in  [80,95]~\mu s\), 
in a similar fashion as in section~\ref{ss:muonN_selection}. 

The neutron-detection efficiency is defined as the ratio of the 
number of detected and incident neutrons.  
The number of incident neutrons is estimated as the count of 
alpha triggers that pass the alpha energy cut. The number of detected 
neutrons is the count of background-subtracted pairs following the 
selected alpha triggers, and with \(t_1 \in T_{\text{DT}}\), 
$\Delta t \in \mathcal{T}$ and $E_2 \in \mathcal{E}_{\text{DT}}$.
The right panel of Fig.~\ref{f:alphaPixels} shows the efficiency 
computed for each alpha sensor pixel, where 
the red rectangle delineates the pixels whose projections fall on the MARS face. 
The inside pixels next to MARS' vertical edges have highly suppressed efficiency 
values due to a higher chance for the incident neutrons to escape the detector 
before being captured and for the Gd gammas to escape if the neutron is captured, 
but also due to the incomplete overlap of the neutron cones corresponding to 
the near-edge pixels. A careful treatment of these data, to be presented in future 
work, will determine the partial overlap of the edge neutron cones and will
accordingly correct the number of incident neutrons. 

In order to provide an estimate of MARS detection efficiency, 
we simply average the efficiency of all inside pixels. 
The systematic uncertainty due to the efficiency non-uniformity is estimated as the 
standard deviation of the inside pixels' efficiency values, 
and represents the dominant contribution to the overall uncertainty. 
Contributions from the \(t_1\) and \(E_2\) cuts are smaller by two 
orders of magnitude. This results in an efficiency to detect 14~MeV 
neutrons equal to (4.22 $\pm$ 1.82)\%.
Furthermore, to obtain an estimate of the incident BRN flux 
from the measured rate, we employ this efficiency value as 
an approximation of the average capture-gated detection efficiency. 
We estimate \(1.20 \pm 0.56~\text{BRN}/\text{m}^2/\text{MWh}\) 
incident on MARS at the center of the Neutrino Alley. 

The proper approach to unfold the incident BRN flux is to model and 
validate MARS response as a function of energy. 
Using a preliminary Geant4 
model of the MARS capture-gated signature that closely
matches the 14~MeV neutron detection efficiency at the center 
of the detector, and assuming 
that the incident BRN flux follows a power-law energy dependence with 
exponent $\alpha = -1.5$ taken from the BRN flux study in~\cite{2017PhDGraysonRich}, 
we calculate that  
the 14~MeV detection efficiency could be overestimating 
the flux-weighted average efficiency by about 50\%. 
Our reported 14~MeV efficiency value already contains an error of that 
magnitude, mainly due to the inaccurate treatment of the near-edge 
pixel data. Future detector modeling and experimental validation work should reduce 
the BRN flux uncertainty due to the detection efficiency and 
will be the subject of a separate publication.


\subsection{\label{ss:fluxComparison} Flux estimation from other detectors}

%


Table~\ref{t:alleyBRNfluxes} summarizes the BRN fluxes measured 
by several neutron detectors along the SNS basement corridor 
per megawatt-hour (MWh) of beam energy.  
A 2015-2016 measurement of the beam neutrons at a neighboring neutrino 
alley location with a scintillator-based neutron double-scatter 
camera (NSC), depicted in Fig.~\ref{f:neutrinoAlley},
produced a flux estimate of \(5.3~\text{beam neutrons}/ \text{m}^2 / \text{MWh}\) 
above 5~MeV within a $1.3~\mu s$ beam window~\cite{CabreraPalmer2019}. 
The uncertainty on the NSC flux value is estimated to be larger than 50\%, 
due to the NSC data's poor statistics (only four double-scatter counts 
were recorded in about five beam-on months) and the significant uncertainty on the 
instrument's poor double-scatter detection efficiency. 
A 2014~NSC measurement in the alcove at the end of the Neutrino Alley, 
also represented in Fig.~\ref{f:neutrinoAlley}, produced 
\( 272.1~\text{beam neutrons}/\text{m}^2 / \text{MWh}\), 
\textit{i.e.,} an almost two orders of magnitude larger flux. 
The 2015~data collected with the SciBath detector~\cite{Heath:2019jpj},
also at the Neutrino Alley alcove, resulted in 
\(13.3~\text{beam neutrons}/\text{m}^2 / \text{MWh}\)
above 5~MeV within the $1~\mu s$ beam window used in that analysis. 
Despite the variation among the flux estimations by the different detectors, 
they consistently indicate a much larger BRN flux at the alcove 
compared to a quieter BRN flux at the Neutrino Alley center.  
MARS operation in the alcove area, which is already in progress, will 
be essential in quantizing the differences in the BRN intensity and spectrum 
between the alcove and the Neutrino Alley center. 


\begin{table}
\centering
\caption{BRN fluxes measured by several neutron detectors in the SNS 
basement corridor. These measurements indicate that the BRN flux in the 
center of the Neutrino Alley is largely suppressed compared to 
the BRN flux reaching the alcove.   
}
\label{t:alleyBRNfluxes}
\begin{tabular}{|c|c|c|c|c|}
\hline
Detector & Neutrino Alley & BRN flux & Uncertainty\\
& location, see Fig.~\ref{f:neutrinoAlley} &\(\text{n}/\text{m}^2/\text{MWh}\) & \%\\
\hline
MARS & center & $1.2$ & 47\\
NSC & center left & $5$ & > 50\\
NSC & alcove wedge & $3\times10^{2}$ & > 50\\
SciBath & alcove wall & $1.3\times10$  & 19\\

\hline
\end{tabular}
\end{table}



\section{\label{s:conclusions}Conclusions}

We have described the analysis and results of the first dataset collected 
with the MARS detector at the SNS. We showed that the intense 511~keV 
gamma-ray background present during beam-on periods has the effect of 
not only increasing the overall event rate but also shifting the 
MARS energy spectrum to the right of the energy axes. We presented a method to 
offset the variable spectral shifts by adjusting the energy cuts so that 
the rate of muon-induced neutron events is kept at the same reference 
value for any data with the same cosmic overburden, where 
the reference rate is computed from five beam-off weeks from April 1 to May 6, 2018.  

Our results show that the MARS detector is able to monitor the low BRN
flux at the center of the Neutrino Alley under nominal SNS operations. 
The measured prompt spectrum decreases with deposited energy up to \(8\)~MeVee, 
which corresponds to $\sim 15$~MeV of incoming neutron energy. 
To properly unfold the BRN spectral flux, the detection-efficiency energy 
dependence must be used. In that regard, the modeling and experimental validation of MARS 
response as a function of incoming neutron energy will the subject of a separate 
publication. 
However, our preliminary simulations show that the fraction of purely elastic 
capture-gated neutron events substantially declines with neutron energy, 
indicting that this detection mode will not provide meaningful bounds on the 
BRN spectra above a few tens of MeV.
The full 22-week exposure of this dataset yields a 17\% statistical 
uncertainty on the total BRN count, while a 2-week integration results in 
a mean of 14 BRN counts per 2-week interval with 58\% statistical uncertainty. 
Sensitivity to the 7\% difference in the operational beam power between the 
two 2018 beam periods (1.3~MW from May to June and 1.4~MW from September to November) 
would have required about 2.6 years of exposure under the signal and background levels 
in the Neutrino Alley center location. 

While the results among different detectors presented in table~\ref{t:alleyBRNfluxes}
vary broadly, they revealed that the BRN flux at the center of the SNS 
basement corridor is suppressed by at least one order of magnitude compared 
to the alcove's BRN flux. To understand the origin of this excess, 
we are examining the possible paths for neutrons from the SNS target 
to Neutrino Alley, and are assessing the amount and location of bulk 
materials attenuating this flux.
Further dedicated BRN measurements in the basement alcove are underway, 
including operating MARS in this location. Under a BRN flux at 
least one order of magnitude larger ---and assuming a background 
rate as in the basement corridor center--- less than a week of MARS data collection 
in the alcove would suffice to achieve the same 17\% statistical uncertainty 
on a BRN count measurement.  

\acknowledgments

The COHERENT collaboration acknowledges the resources generously provided by the 
Spallation Neutron Source, a DOE Office of Science User Facility operated by the 
Oak Ridge National Laboratory. This work was supported by the US Department of 
Energy (DOE), Office of Science, Office of High Energy Physics and Office of Nuclear 
Physics; the National Science Foundation; the Consortium for Nonproliferation Enabling 
Capabilities; the Institute for Basic Science (Korea, grant no.\,IBS-R017-G1-2019-a00); 
the Ministry of Science and Higher Education of the Russian Federation (Project 
``Fundamental properties of elementary particles and cosmology'' No. 0723-2020-0041); 
the Russian Foundation for Basic Research (Project 20-02-00670\_a); and the US DOE 
Office of Science Graduate Student Research (SCGSR) program, administered for DOE 
by the Oak Ridge Institute for Science and Education which is in turn managed by 
Oak Ridge Associated Universities. Sandia National Laboratories is a multi-mission 
laboratory managed and operated by National Technology and Engineering Solutions 
of Sandia LLC, a wholly owned subsidiary of Honeywell International Inc., for the 
U.S. Department of Energy's National Nuclear Security Administration under contract 
DE-NA0003525. The Triangle Universities Nuclear Laboratory is supported by the U.S. 
Department of Energy under grant DE-FG02-97ER41033. Laboratory Directed Research 
and Development funds from Oak Ridge National Laboratory also supported this project. 
This research used the Oak Ridge Leadership Computing Facility, which is a DOE Office 
of Science User Facility.

We acknowledge the NMDB database (www.nmdb.eu), founded under the European Union's FP7 
programme (contract no. 213007) for providing data. The neutron monitor data from 
Newark/Swarthmore are provided by the University of Delaware Department of Physics and 
Astronomy and the Bartol Research Institute.

\bibliography{mars_sns_bibfile}

\end{document}

%% file: authors/authors-MARS2018.tex
\newcommand{\Mephi}{a}
\newcommand{\Mephidesc}{\affiliation[\Mephi]{National Research Nuclear University MEPhI (Moscow Engineering Physics Institute), Moscow, 115409, Russian Federation}}
\newcommand{\Duke}{b}
\newcommand{\Dukedesc}{\affiliation[\Duke]{Department of Physics, Duke University, Durham, NC, 27708, USA}}
\newcommand{\TUNL}{c}
\newcommand{\TUNLdesc}{\affiliation[\TUNL]{Triangle Universities Nuclear Laboratory, Durham, NC, 27708, USA}}
\newcommand{\UTK}{d}
\newcommand{\UTKdesc}{\affiliation[\UTK]{Department of Physics and Astronomy, University of Tennessee, Knoxville, TN, 37996, USA}}
\newcommand{\ITEP}{e}
\newcommand{\ITEPdesc}{\affiliation[\ITEP]{Institute for Theoretical and Experimental Physics named by A.I. Alikhanov of National Research Centre ``Kurchatov Institute'', Moscow, 117218, Russian Federation}}
\newcommand{\ORNL}{f}
\newcommand{\ORNLdesc}{\affiliation[\ORNL]{Oak Ridge National Laboratory, Oak Ridge, TN, 37831, USA}}
\newcommand{\USD}{g}
\newcommand{\USDdesc}{\affiliation[\USD]{Physics Department, University of South Dakota, Vermillion, SD, 57069, USA}}
\newcommand{\NCSU}{h}
\newcommand{\NCSUdesc}{\affiliation[\NCSU]{Department of Physics, North Carolina State University, Raleigh, NC, 27695, USA}}
\newcommand{\Sandia}{i}
\newcommand{\Sandiadesc}{\affiliation[\Sandia]{Sandia National Laboratories, Livermore, CA, 94550, USA}}
\newcommand{\UW}{j}
\newcommand{\UWdesc}{\affiliation[\UW]{Center for Experimental Nuclear Physics and Astrophysics \& Department of Physics, University of Washington, Seattle, WA, 98195, USA}}
\newcommand{\LANL}{k}
\newcommand{\LANLdesc}{\affiliation[\LANL]{Los Alamos National Laboratory, Los Alamos, NM, 87545, USA}}
\newcommand{\Laurentian}{l}
\newcommand{\Laurentiandesc}{\affiliation[\Laurentian]{Department of Physics, Laurentian University, Sudbury, Ontario, P3E 2C6, Canada}}
\newcommand{\CMU}{m}
\newcommand{\CMUdesc}{\affiliation[\CMU]{Department of Physics, Carnegie Mellon University, Pittsburgh, PA, 15213, USA}}
\newcommand{\IU}{n}
\newcommand{\IUdesc}{\affiliation[\IU]{Department of Physics, Indiana University, Bloomington, IN, 47405, USA}}
\newcommand{\VT}{o}
\newcommand{\VTdesc}{\affiliation[\VT]{Center for Neutrino Physics, Virginia Tech, Blacksburg, VA, 24061, USA}}
\newcommand{\NCCU}{p}
\newcommand{\NCCUdesc}{\affiliation[\NCCU]{Department of Mathematics and Physics, North Carolina Central University, Durham, NC, 27707, USA}}
\newcommand{\UF}{q}
\newcommand{\UFdesc}{\affiliation[\UF]{Department of Physics, University of Florida, Gainesville, FL, 32611, USA}}
\newcommand{\Tufts}{r}
\newcommand{\Tuftsdesc}{\affiliation[\Tufts]{Department of Physics and Astronomy, Tufts University, Medford, MA, 02155, USA}}
\newcommand{\SNU}{s}
\newcommand{\SNUdesc}{\affiliation[\SNU]{Department of Physics and Astronomy, Seoul National University, Seoul, 08826, Korea}}
\author[\Mephi]{D.~Akimov,}\Mephidesc
\author[\Duke,\TUNL]{P.~An,}\Dukedesc\TUNLdesc
\author[\Duke,\TUNL]{C.~Awe,}
\author[\Duke,\TUNL]{P.S.~Barbeau,}
\author[\UTK]{B.~Becker,}\UTKdesc
\author[\ITEP,\Mephi]{V.~Belov ,}\ITEPdesc
\author[\UTK]{I.~Bernardi,}
\author[\ORNL]{M.A.~Blackston,}\ORNLdesc
\author[\USD]{C.~Bock,}\USDdesc
\author[\Mephi]{A.~Bolozdynya,}
\author[\NCSU]{J.~Browning,}\NCSUdesc
\author[\Sandia]{B.~Cabrera-Palmer,}\Sandiadesc
\author[\USD,1]{D.~Chernyak,}\note{Now at:  Moscow, "Theoretical and Experimental Physics,  117218,  Russia", Department of Physics and Astronomy, Tuscaloosa and Institute for Nuclear Research of NASU, Kyiv, 03028, Ukraine}
\author[\Duke]{E.~Conley,}
\author[\ORNL]{J.~Daughhetee,}
\author[\UW]{J.~Detwiler,}\UWdesc
\author[\USD]{K.~Ding,}
\author[\UW]{M.R.~Durand,}
\author[\UTK,\ORNL]{Y.~Efremenko,}
\author[\LANL]{S.R.~Elliott,}\LANLdesc
\author[\ORNL]{L.~Fabris,}
\author[\ORNL]{M.~Febbraro,}
\author[\Laurentian]{A.~Gallo Rosso,}\Laurentiandesc
\author[\ORNL,\UTK]{A.~Galindo-Uribarri,}
\author[\TUNL,\ORNL,\NCSU]{M.P.~Green ,}
\author[\ORNL]{M.R.~Heath,}
\author[\Duke,\TUNL]{S.~Hedges,}
\author[\CMU]{D.~Hoang,}\CMUdesc
\author[\IU]{M.~Hughes,}\IUdesc
\author[\IU]{B.A.~Johnson,}
\author[\Duke,\TUNL]{T.~Johnson,}
\author[\Mephi]{A.~Khromov,}
\author[\Mephi,\ITEP]{A.~Konovalov,}
\author[\Mephi,\ITEP]{E.~Kozlova,}
\author[\Mephi]{A.~Kumpan,}
\author[\Duke,\TUNL]{L.~Li,}
\author[\VT]{J.M.~Link,}\VTdesc
\author[\USD]{J.~Liu,}
\author[\NCSU]{K.~Mann,}
\author[\NCCU,\TUNL]{D.M.~Markoff,}\NCCUdesc
\author[\IU]{J.~Mastroberti,}
\author[\ORNL]{P.E.~Mueller,}
\author[\ORNL]{J.~Newby,}
\author[\CMU]{D.S.~Parno,}
\author[\ORNL]{S.I.~Penttila,}
\author[\Duke]{D.~Pershey,}
\author[\CMU]{R.~Rapp,}
\author[\UF]{H.~Ray,}\UFdesc
\author[\Duke]{J.~Raybern,}
\author[\Mephi,\ITEP]{O.~Razuvaeva,}
\author[\Sandia]{D.~Reyna,}
\author[\TUNL]{G.C.~Rich,}
\author[\NCCU,\TUNL]{J.~Ross,}
\author[\Mephi]{D.~Rudik,}
\author[\Duke,\TUNL]{J.~Runge,}
\author[\IU]{D.J.~Salvat,}
\author[\CMU]{A.M.~Salyapongse,}
\author[\USD]{J.~Sander,}
\author[\Duke]{K.~Scholberg,}
\author[\Mephi]{A.~Shakirov,}
\author[\Mephi,\ITEP]{G.~Simakov,}
\author[\Duke,2]{G.~Sinev,}\note{Now at: South Dakota School of Mines and Technology, Rapid City, SD, 57701, USA}
\author[\IU]{W.M.~Snow,}
\author[\Mephi]{V.~Sosnovstsev,}
\author[\IU]{B.~Suh,}
\author[\IU]{R.~Tayloe,}
\author[\VT]{K.~Tellez-Giron-Flores,}
\author[\IU,3]{I.~Tolstukhin,}\note{Now at: Argonne National Laboratory, Argonne, IL, 60439, USA}
\author[\NCCU,\TUNL]{E.~Ujah,}
\author[\IU]{J.~Vanderwerp,}
\author[\ORNL]{R.L.~Varner,}
\author[\Laurentian]{C.J.~Virtue,}
\author[\IU]{G.~Visser,}
\author[\Tufts]{T.~Wongjirad,}\Tuftsdesc
\author[\CMU]{Y.-R.~Yen,}
\author[\SNU]{J.~Yoo,}\SNUdesc
\author[\ORNL]{C.-H.~Yu,}
\author[\IU,4]{J.~Zettlemoyer}\note{Now at: Fermi National Accelerator Laboratory, Batavia, IL, 60510, USA}

%% file: BeamNeutronResult2018Data.bbl
\providecommand{\noopsort}[1]{}\providecommand{\singleletter}[1]{#1}%

\providecommand{\href}[2]{#2}\begingroup\raggedright\begin{thebibliography}{10}

\bibitem{COHERENT:2017ipa}
{\scshape COHERENT} collaboration, D.~Akimov, J.~B. Albert, P.~An, C.~Awe,
  P.~S. Barbeau, B.~Becker et~al., \emph{{Observation of Coherent Elastic
  Neutrino-Nucleus Scattering}},
  \href{http://dx.doi.org/10.1126/science.aao0990}{\emph{Science} {\bfseries
  357} (2017) 1123--1126}, [\href{https://arxiv.org/abs/1708.01294}{{\ttfamily
  1708.01294}}].

\bibitem{COHERENT:2018imc}
{\scshape COHERENT} collaboration, D.~Akimov et~al., \emph{{COHERENT
  Collaboration data release from the first observation of coherent elastic
  neutrino-nucleus scattering}},
  \href{https://arxiv.org/abs/1804.09459}{{\ttfamily 1804.09459}}.

\bibitem{COHERENT:2018gft}
{\scshape COHERENT} collaboration, D.~Akimov, J.~B. Albert, P.~An, C.~Awe,
  P.~S. Barbeau, B.~Becker et~al., \emph{{COHERENT 2018 at the Spallation
  Neutron Source}},  \href{https://arxiv.org/abs/1803.09183}{{\ttfamily
  1803.09183}}.

\bibitem{COHERENT:2019iyj}
{\scshape COHERENT} collaboration, D.~Akimov et~al., \emph{{First Constraint on
  Coherent Elastic Neutrino-Nucleus Scattering in Argon}},
  \href{http://dx.doi.org/10.1103/PhysRevD.100.115020}{\emph{Phys. Rev. D}
  {\bfseries 100} (2019) 115020},
  [\href{https://arxiv.org/abs/1909.05913}{{\ttfamily 1909.05913}}].

\bibitem{COHERENT:2020iec}
{\scshape COHERENT} collaboration, D.~Akimov et~al., \emph{{First Measurement
  of Coherent Elastic Neutrino-Nucleus Scattering on Argon}},
  \href{http://dx.doi.org/10.1103/PhysRevLett.126.012002}{\emph{Phys. Rev.
  Lett.} {\bfseries 126} (2021) 012002},
  [\href{https://arxiv.org/abs/2003.10630}{{\ttfamily 2003.10630}}].

\bibitem{COHERENT:2020ybo}
{\scshape COHERENT} collaboration, D.~Akimov et~al., \emph{{COHERENT
  Collaboration data release from the first detection of coherent elastic
  neutrino-nucleus scattering on argon}},
  \href{https://arxiv.org/abs/2006.12659}{{\ttfamily 2006.12659}}.

\bibitem{COHERENT:2021xhx}
{\scshape COHERENT} collaboration, D.~Akimov et~al., \emph{{A D$_{2}$O detector
  for flux normalization of a pion decay-at-rest neutrino source}},
  \href{https://arxiv.org/abs/2104.09605}{{\ttfamily 2104.09605}}.

\bibitem{COHERENT:2019kwz}
{\scshape COHERENT} collaboration, D.~Akimov et~al., \emph{{Sensitivity of the
  COHERENT Experiment to Accelerator-Produced Dark Matter}},
  \href{http://dx.doi.org/10.1103/PhysRevD.102.052007}{\emph{Phys. Rev. D}
  {\bfseries 102} (2020) 052007},
  [\href{https://arxiv.org/abs/1911.06422}{{\ttfamily 1911.06422}}].

\bibitem{collaboration2015coherent}
{\scshape COHERENT} collaboration, D.~Akimov, P.~An, C.~Awe, P.~S. Barbeau,
  P.~Barton, B.~Becker et~al., \emph{{The COHERENT Experiment at the Spallation
  Neutron Source}},  \href{https://arxiv.org/abs/1509.08702}{{\ttfamily
  1509.08702}}.

\bibitem{ROECKER201621}
C.~Roecker, A.~Bernstein, N.~Bowden, B.~Cabrera-Palmer, S.~Dazeley, M.~Gerling
  et~al., \emph{{Design of a transportable high efficiency fast neutron
  spectrometer}},
  \href{http://dx.doi.org/https://doi.org/10.1016/j.nima.2016.04.032}{\emph{Nuclear
  Instruments and Methods in Physics Research Section A: Accelerators,
  Spectrometers, Detectors and Associated Equipment} {\bfseries 826} (2016) 21
  -- 30}.

\bibitem{geant4}
{\scshape GEANT4} collaboration, S.~Agostinelli et~al., \emph{{GEANT4: A
  simulation toolkit}},
  \href{http://dx.doi.org/10.1016/S0168-9002(03)01368-8}{\emph{NIM} {\bfseries
  A506} (2003) 250--303}.

\bibitem{Struck}
``{SIS3316 16 channel VME digitizer family}.'' [Online]
  http://www.struck.de/sis3316.html.

\bibitem{NGM}
``{A toolkit for data acquisition control and analysis for waveform digitizers
  including GAGE, XIA, and Struck}.'' [Online]
  https://code.ornl.gov/CASA/ngmdaq.

\bibitem{CabreraPalmer2019}
B.~Cabrera-Palmer, E.~Brubaker, M.~Gerling and D.~Reyna, \emph{{Extension of
  the neutron scatter camera sensitivity to the 10-200 MeV neutron energy
  range}}, \href{http://dx.doi.org/10.1063/1.5091715}{\emph{Review of
  Scientific Instruments} {\bfseries 90} (2019) 053305},
  [\href{https://arxiv.org/abs/https://doi.org/10.1063/1.5091715}{{\ttfamily
  https://doi.org/10.1063/1.5091715}}].

\bibitem{Brice:2013fwa}
S.~Brice et~al., \emph{A method for measuring coherent elastic neutrino-nucleus
  scattering at a far off-axis high-energy neutrino beam target},
  \href{http://dx.doi.org/10.1103/PhysRevD.89.072004}{\emph{Phys. Rev. D}
  {\bfseries 89} (2014) 072004},
  [\href{https://arxiv.org/abs/1311.5958}{{\ttfamily 1311.5958}}].

\bibitem{roecker_2016}
C.~D. Roecker, \emph{{Measurement of the High-Energy Neutron Flux Above and
  Below Ground}}.
\newblock PhD thesis, U.C. Berkeley, 2016.

\bibitem{Web:Grafana:Docs}
{Grafana Labs}, ``{Grafana Documentation}.'' [Online]
  https://grafana.com/docs/, 2018.

\bibitem{Jian_Fu_2010}
Z.~Jian-Fu, R.~Xi-Chao, H.~Long, L.~Xia, B.~Jie, Z.~Guo-Guang et~al.,
  \emph{{Measurements of the light output functions of plastic scintillator
  using $^9$Be(d, n)$^{10}$B reaction neutron source}},
  \href{http://dx.doi.org/10.1088/1674-1137/34/7/011}{\emph{Chinese Physics C}
  {\bfseries 34} (jul, 2010) 988--992}.

\bibitem{Laplace:2020mfy}
T.~Laplace et~al., \emph{{Low Energy Light Yield of Fast Plastic
  Scintillators}},
  \href{http://dx.doi.org/10.1016/j.nima.2018.10.122}{\emph{Nucl. Instrum.
  Meth. A} {\bfseries 954} (2020) 161444},
  [\href{https://arxiv.org/abs/2009.07217}{{\ttfamily 2009.07217}}].

\bibitem{Agashe:2014kda}
{\scshape Particle Data Group} collaboration, K.~A. Olive et~al., \emph{{Review
  of Particle Physics}},
  \href{http://dx.doi.org/10.1088/1674-1137/38/9/090001}{\emph{Chin. Phys.}
  {\bfseries C38} (2014) 090001}.

\bibitem{Acciarri:2017sjy}
{\scshape MicroBooNE} collaboration, R.~Acciarri et~al., \emph{{Michel Electron
  Reconstruction Using Cosmic-Ray Data from the MicroBooNE LArTPC}},
  \href{http://dx.doi.org/10.1088/1748-0221/12/09/P09014}{\emph{JINST}
  {\bfseries 12} (2017) P09014},
  [\href{https://arxiv.org/abs/1704.02927}{{\ttfamily 1704.02927}}].

\bibitem{Hertenberger:1995ae}
R.~Hertenberger, M.~Chen and B.~L. Dougherty, \emph{{Muon induced neutron and
  pion production in an organic liquid scintillator at a shallow depth}},
  \href{http://dx.doi.org/10.1103/PhysRevC.52.3449}{\emph{Phys. Rev. C}
  {\bfseries 52} (1995) 3449--3459}.

\bibitem{Boehm:2000ru}
F.~Boehm et~al., \emph{{Neutron production by cosmic ray muons at shallow
  depth}}, \href{http://dx.doi.org/10.1103/PhysRevD.62.092005}{\emph{Phys. Rev.
  D} {\bfseries 62} (2000) 092005},
  [\href{https://arxiv.org/abs/hep-ex/0006014}{{\ttfamily hep-ex/0006014}}].

\bibitem{PhysRevD.64.013012}
Y.-F. Wang, V.~Balic, G.~Gratta, A.~Fass\`o, S.~Roesler and A.~Ferrari,
  \emph{Predicting neutron production from cosmic-ray muons},
  \href{http://dx.doi.org/10.1103/PhysRevD.64.013012}{\emph{Phys. Rev. D}
  {\bfseries 64} (Jun, 2001) 013012}.

\bibitem{osti_1151758}
L.~Garrison, \emph{Measurement of Neutron and Muon Fluxes 100~m Underground
  with the SciBath Detector}.
\newblock PhD thesis, Jan, 2014.
\newblock 10.2172/1151758.

\bibitem{Clem2000}
J.~M. Clem and L.~I. Dorman, \emph{Neutron monitor response functions},
  \href{http://dx.doi.org/10.1023/A:1026508915269}{\emph{Space Science Reviews}
  {\bfseries 93} (2000) 335--359}.

\bibitem{NMDB}
``{Newark/Swarthmore Neutron Monitor}.'' [Online] https://www.nmdb.eu/.

\bibitem{BOWDEN2012209}
N.~Bowden, M.~Sweany and S.~Dazeley, \emph{{A note on neutron capture
  correlation signals, backgrounds, and efficiencies}},
  \href{http://dx.doi.org/https://doi.org/10.1016/j.nima.2012.07.005}{\emph{Nuclear
  Instruments and Methods in Physics Research Section A: Accelerators,
  Spectrometers, Detectors and Associated Equipment} {\bfseries 693} (2012) 209
  -- 214}.

\bibitem{API120}
``{Thermo Fisher Scientific API 120 Neutron Generator}.'' [Online]
  https://www.thermofisher.com/.

\bibitem{CHICHESTER2005753}
D.~Chichester, M.~Lemchak and J.~Simpson, \emph{The api 120: A portable neutron
  generator for the associated particle technique},
  \href{http://dx.doi.org/https://doi.org/10.1016/j.nimb.2005.07.128}{\emph{Nuclear
  Instruments and Methods in Physics Research Section B: Beam Interactions with
  Materials and Atoms} {\bfseries 241} (2005) 753--758}.

\bibitem{Beyerle1990}
A.~Beyerle, J.~P. Hurley and L.~Tunnell, \emph{Design of an associated particle
  imaging system}, {\emph{Nuclear Inst.and Methods in Physics Research, A}
  {\bfseries 299} (1990) 458--462}.

\bibitem{osti_4036016}
H.~O. Anger, \emph{Scintillation camera with multichannel collimators},
  {\emph{Journal of Nuclear Medicine (U.S.)} {\bfseries 5} (7, 1964) }.

\bibitem{2017PhDGraysonRich}
G.~C. {Rich}, \emph{{Measurement of Low-Energy Nuclear-Recoil Quenching Factors
  in CsI[Na] and Statistical Analysis of the First Observation of Coherent,
  Elastic Neutrino-Nucleus Scattering}}.
\newblock PhD thesis, The University of North Carolina at Chapel Hill, Jan.,
  2017.

\bibitem{Heath:2019jpj}
M.~R. Heath, \emph{{A First Search for Coherent Elastic Neutrino-Nucleus
  Scattering with Liquid Argon}}.
\newblock PhD thesis, Indiana U., Bloomington (main), 2019.
\newblock 10.5967/jmrj-9078.

\end{thebibliography}\endgroup
